\documentclass{llncs}
\usepackage{graphicx} 
\usepackage{epstopdf} 
\usepackage{amsmath} 
\usepackage{amsmath,bm} 
\usepackage[noend]{algpseudocode}  
\usepackage{algorithmicx,algorithm} 
\usepackage{multirow} 
\usepackage{makecell,multirow,diagbox} 
\usepackage{subfigure} 
\usepackage{array} 
\usepackage{amsfonts} 
\usepackage{float} %
\usepackage{bbding}
\usepackage[misc]{ifsym}
\makeatletter
\newcolumntype{"}{@{\hskip\tabcolsep\vrule width 1pt\hskip\tabcolsep}}
\makeatother
\begin{document}
\title{Cross-Domain Recommendation for Cold-Start Users via Neighborhood Based Feature Mapping}
\author{Xinghua Wang\inst{1} \and Zhaohui Peng\inst{1}$^{\left(\text{\Letter}\right)}$ \and Senzhang Wang\inst{2} \and Philip S. Yu \inst{3 \and 4} \and Wenjing Fu\inst{1} \and Xiaoguang Hong\inst{1}}
\institute{School of Computer Science and Technology,\\ Shandong University, Jinan, China\\ \email{wang.xingh@foxmail.com, \{pzh,hxg\}@sdu.edu.cn,\\fuwenjing@mail.sdu.edu.cn} \and College of Computer Science and Technology, Nanjing University\\ of Aeronautics and Astronautics, Nanjing, China\\ \email{szwang@nuaa.edu.cn} \and Department of Computer Science,\\ University of Illinois at Chicago, Chicago, USA\\ \email{psyu@uic.edu}\\ \and Institute for Data Science, Tsinghua University, Beijing, China}
\maketitle
\begin{abstract}
Collaborative Filtering (CF) is a widely adopted technique in recommender systems. Traditional CF models mainly focus on predicting a user's preference to the items in a single domain such as the movie domain or the music domain. A major challenge for such models is the data sparsity problem, and especially, CF cannot make accurate predictions for the cold-start users who have no ratings at all. Although Cross-Domain Collaborative Filtering (CDCF) is proposed for effectively transferring users' rating preference across different domains, it is still difficult for existing CDCF models to tackle the cold-start users in the target domain due to the extreme data sparsity. In this paper, we propose a Cross-Domain Latent Feature Mapping (CDLFM) model for cold-start users in the target domain. Firstly, in order to better characterize users in sparse domains, we take the users' similarity relationship on rating behaviors into consideration and propose the Matrix Factorization by incorporating User Similarities (MFUS) in which three similarity measures are proposed. Next, to perform knowledge transfer across domains, we propose a neighborhood based gradient boosting trees method to learn the cross-domain user latent feature mapping function. For each cold-start user, we learn his/her feature mapping function based on the latent feature pairs of those linked users who have similar rating behaviors with the cold-start user in the auxiliary domain. And the preference of the cold-start user in the target domain can be predicted based on the mapping function and his/her latent features in the auxiliary domain. Experimental results on two real datasets extracted from Amazon transaction data demonstrate the superiority of our proposed model against other state-of-the-art methods.
\begin{keywords}
cross-domain recommendation, cold start, feature mapping
\end{keywords}
\end{abstract}
\section{Introduction}
With the quick development of Internet and Web techniques, e-commerce has become increasingly popular and greatly changed people's purchasing behaviors. Shopping online can provide users with diversified choices by which users are more likely to be overwhelmed. In order to help consumers find what they really desire from the massive amounts of products, recommender systems become indispensable in most e-commerce websites.

Collaborative Filtering (CF) is a widely used technique in recommender systems due to the fact that it requires little domain-specific knowledge, and yet it can address data aspects that are often difficult to profile using content filtering \cite{MF}. Traditional CF models focus on single-domain user preference prediction and suffer from the data sparsity problem. In fact, there are many item domains and the user preference in different domains are correlated. For example, users who like comedy movies usually prefer humorous books. Therefore, Cross-Domain Collaborative Filtering (CDCF) is proposed to enrich the knowledge in the target domain by taking advantage of multi-domain ratings and becomes an emerging research topic. Even so, it is still very challenging to make reliable recommendations for the cold-start users in one domain due to the extreme data sparsity. And most CDCF models, e.g. CBT\cite{Coodbook}, RMGM \cite{RMGM}, TCF \cite{Transfer in CF}, are designed to alleviate the single-domain data sparsity problem, while how to effectively recommend for the cold-start users is still not fully explored.

\begin{figure} [t]
\centering
\includegraphics[width=12.2cm]{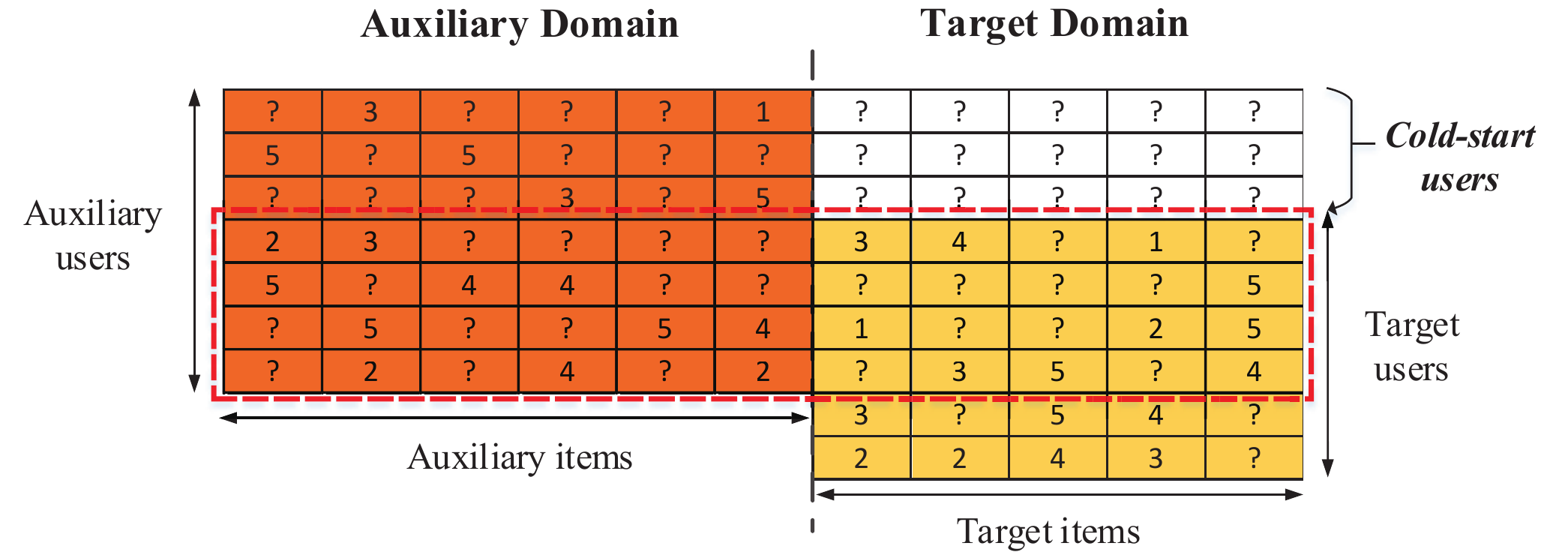}
\caption{Illustration of the cross-domain recommendation for cold-start users}
\label{f1}
\end{figure}

In real world, the cold-start users of an item domain may have ratings in another item domain. For example, in one of the largest Chinese B2C e-commerce website Jingdong\footnote{https://www.jd.com}, most users tend to buy electronic products while are much less interested in purchasing other types of products such as books and foods. On the contrary, in the e-commerce website Dangdang\footnote{http://www.dangdang.com/}, users prefer to purchase books and will probably not purchase its electronic products. Thus, we focus on studying the cross-domain recommendation for cold-start users and the problem setting studied in this paper is illustrated in Fig. \ref{f1}. We differentiate item domains based on the item type level addressed by the work \cite{Recommender handbook}. For example, movies and books represent different domains, but horror movies and comedy movies belong to the same domain. One can see from Fig. \ref{f1} that in our problem setting cold-start users only have ratings in the auxiliary domain, which is different from most previous works \cite{Coodbook,RMGM,Transfer in CF,Transfer sparsity} that assume the auxiliary domain data is relatively denser than the target domain data without considering cold-start users. Users who have ratings in both domains are called linked users whose rating data is marked with dashed red box in Fig. \ref{f1}. Linked users serve as a bridge for our model to transfer knowledge across domains. It is challenging to make recommendations for the cold-start users in the target domain. First, rating matrices in different item domains are usually sparse, thus how to better model the unique characteristics of users in different domains becomes very important. Second, there is no rating data for cold-start users in the target domain and user rating behaviors or preference in different domains are correlated but different. Therefore, what knowledge should be transferred and how to transfer the knowledge across domains remain an open problem.

To address the above challenges, we propose a Cross-Domain Latent Feature Mapping (CDLFM) model. Firstly, we handle the rating matrices in different domains separately with Matrix Factorization by incorporating User Similarities (MFUS) in order to gain domain-specific user latent features in sparse domains. For better characterising users, besides the observed ratings, we also take users' other rating behaviors into consideration, such as unrated products and rating biases. We first compute the rating behavior based user similarities via three proposed similarity measures and then embed the user similarities into the matrix factorization model. Next, to transfer the knowledge of user characteristics across domains, we propose a neighborhood based gradient boosting trees method to learn the cross-domain user latent feature mapping function. For each cold-start user, we first find his/her nearest linked users, with whom he/she has similar rating behaviors in the auxiliary domain, and then the latent feature pairs of these linked users are used to learn the feature mapping function. According to the learned mapping function and the cold-start user's latent features in the auxiliary domain, we predict his/her characteristics in the target domain and make the preference prediction. Our major contributions are summarized as follows:
\begin{itemize}
\item[$\bullet$] An improved rating matrix factorization model is proposed which is the first to consider users' similarity relationship reflected from their rating behaviors.
\item[$\bullet$] A neighborhood based gradient boosting trees method is proposed for more accurately performing cross-domain latent feature mapping.
\item[$\bullet$] We conduct extensive experiments on the Amazon rating data to evaluate the proposed model and make comparisons with other state-of-the-art models.
\end{itemize}
\section{Related Work}
In this section, we discuss the existing research related to our work, mainly including rating matrix factorization and cross-domain recommendation.

There have been a lot of models \cite{MF,Probabilistic,Informative,Factorizationlibfm,FactorizationNeighborhood,TagiCoFi} for single-domain rating matrix factorization aiming to fit the observed ratings more accurately and effectively. Koren \cite{FactorizationNeighborhood} argues that neighborhood models can effectively detect localized relationships and latent factor models are generally effective at estimating overall structure. Our MFUS capitalizes on the advantages of both the two methods by incorporating user similarities into the matrix factorization process. Unlike Koren, we do not introduce extra parameters. TagiCoFi \cite{TagiCoFi} also aims to improve the performance of traditional Matrix Factorization (MF) model \cite{Probabilistic}, but it relies on tagging information.

For cross-domain recommendation \cite{Recommender handbook,Cross-domain systems}, transfer learning \cite{a survey of TL,A survey of tl with auxidata,CDKTL} has been used extensively for alleviating the data sparsity problem \cite{Coodbook,RMGM,Transfer in CF,Transfer sparsity}. CMF \cite{CMF} couples two matrices on common dimension by sharing the same factor matrix. For rating matrices, CMF only learns one user factor matrix without considering the heterogeneity between the latent features in different domains. The model proposed in \cite{Multi} can be viewed as a generalization of CMF where each domain has its own user factor matrix, but its focus is also on the data sparsity problem without considering the cold-start users. For the cold start problem, there have been tag-based and review-based cross-domain factorization models \cite{Exploiting social tags,Tianhang}. Hu \cite{CDTF} mentions the unacquainted world for users and propose CDTF to capture the triadic relation of user-item-domain by tensor factorization. EMCDR \cite{EMCDR} and \cite{TMatrix} try to use the Multi-Layer Perceptron (MLP) and a transformation matrix to map the user feature vector across domains, but they take all the linked users into consideration which may introduce noise. On social networks \cite{MMRate}, the cold start problem has been widely studied. Zhou \cite{VLDBJ} works on the cold-start batch video recommendation in shared community. Zhao \cite{Connecting socialmedia} aims to recommend products from e-commerce websites to users at social networks in cold-start situations and they map users' social networking features to another feature representation for product recommendation. In our work, no text information is available and we make cross-domain latent feature mapping in a more explicable way.
\section{Problem Formulation}
Two item domains are involved in our model. In the target domain, we have the rating matrix $\mathbf{R}^{t}\in\{1,2,3,4,5,?\}^{\left|U_{t}\right|\times\left|P_{t}\right|}$ where the question mark $"?"$ denotes a missing rating value and $U_{t}$, $P_{t}$ denote the user and product set in the target domain. Likewise, we have $\mathbf{R}^{a}\in\{1,2,3,4,5,?\}^{\left|U_{a}\right|\times\left|P_{a}\right|}$ in the auxiliary domain. We formally define the studied problem as follows.
\begin{definition}
(Cross-domain Recommendation for Cold-start Users)
Given two rating matrices $\mathbf{R}^{t}$ and $\mathbf{R}^{a}$ of two item domains, there are some cold-start users $U_{T}$ who only have ratings in the auxiliary domain while have no ratings in the target domain, and some linked users $U_{L}$ who have ratings in both domains. Our goal is to take $U_{L}$ as a bridge to transfer knowledge from the auxiliary domain to the target domain for predicting the preference of users $U_{T}$ on items $P_{t}$.
\end{definition}

\begin{figure} [t]
\centering
\includegraphics[width=12.2cm]{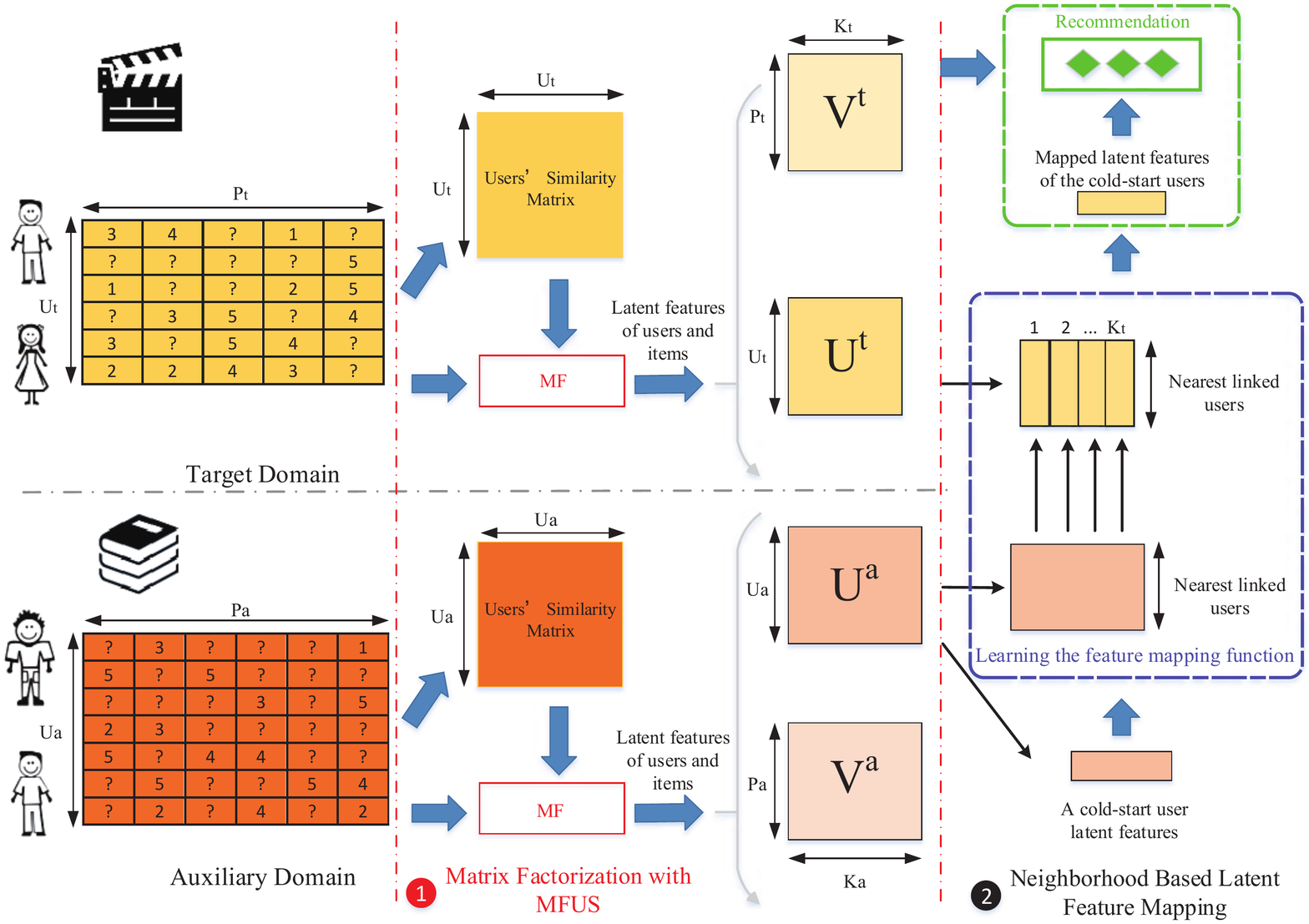}
\caption{The workflow diagram for our proposed model CDLFM}
\label{f2}
\end{figure}

The workflow of our model CDLFM is shown in Fig. \ref{f2} which consists of two major steps. In the first step, we propose an improved rating matrix factorization model to learn the latent features of users and items in the two domains separately. In the second step, we propose a neighborhood based latent feature mapping method to learn the mapping function for each cold-start user. According to the cold-start user's latent features in the auxiliary domain and the learned mapping function, we obtain his/her mapped latent features in the target domain and make recommendations.
\section{Matrix Factorization by Incorporating User Similarities}
The user rating behaviors in different item domains can be quite different. For example, a user may have a lot of ratings for electronic products but have a little for clothes. Another example is that users may use the aspects including acting skills, scenes, lines, etc. to evaluate a movie, while they will use quite different aspects to evaluate a book. Therefore, in the first step of our model, we handle the rating matrices of different domains separately in order to gain domain-specific latent features of users. In order to better characterize users in sparse domains, we take users' rating behaviors into consideration and an improved rating matrix factorization model named MFUS ($\mathbf{M}$atrix $\mathbf{F}$actorization by incorporating $\mathbf{U}$ser $\mathbf{S}$imilarities) is proposed. In MFUS, we first compute the similarities between users based on their rating behaviors and then embed these similarities into the matrix factorization process.
\subsection{Rating Behavior Based User Similarity Measures}
\subsubsection{Similarity Based on Common Ratings.}
Given two users $u$ and $v$, if they have commonly rated products $C_{uv}$, we can compute their similarity based on their rating similarity on $C_{uv}$. We use a matrix $\mathbf{A}^{\left(1\right)}$ to denote their ratings on $C_{uv}$ and the elements $A^{\left(1\right)}_{ui}$ and $A^{\left(1\right)}_{vi}$ represent their ratings on the product $i$. We can compute the first similarity measure between $u$ and $v$ as follows:
\[
D_{uv}^{\left(1\right)}=\sum_{z=1}^{\left|C_{uv}\right|}\left(A_{uz}^{\left(1\right)}-A_{vz}^{\left(1\right)}\right)^{2}, \qquad
S_{uv}^{\left(1\right)}=e^{-\frac{\gamma_{1}D_{uv}^{\left(1\right)}}{\left|C_{uv}\right|}} \; \left(\gamma_{1}>0\right)
\]
Here, we adopt an exponential function to transform users' rating difference into a similarity value and $\gamma_{1}$ is a predefined parameter. $D_{uv}^{\left(1\right)}$ measures the squared difference of their common ratings, and a small $D_{uv}^{\left(1\right)}$ means a large similarity.
\subsubsection{Similarity Based on the Estimations of Having No Interest.}
Besides the rated products, a user's potential preference can be also reflected by the products that he/she does not give ratings to. The reason is that purchase always happens after comparison and evaluation. However, we can not arbitrarily conclude that a user does not like the unrated products.

We use $P_{ui}$ to represent the probability of user $u$ having no interest on the product $i$. If $u$ dose not rate $i$, $P_{ui}$ can be estimated by the following formula:
\begin{equation} \label{e1}
P_{ui}=\left[1-f_{1}\left(n_{u}\right)\times f_{2}\left(n_{i}\right)\right]\times\left[1-f_{3}\left(\frac{n_{i}}{n}\right)\times f_{3}\left(\frac{n_{Hi}}{n_{i}}\right)\right]
\end{equation}
where $f_{1}\left(n_{u}\right)=\sqrt{1-\frac{n_{u}^{2}}{m^{2}}}$, $f_{2}\left(n_{i}\right)=\sqrt{1-\frac{n_{i}^{2}}{n^{2}}}$, $f_{3}\left(x\right)=\frac{2}{1+e^{-\sigma x}}-1$. $n$ and $m$ denote the numbers of users and products in a domain. $n_{u}$ and $n_{i}$ represent the total rating numbers of user $u$ and product $i$. $n_{Hi}$ is the number of high ratings on product $i$ (the high rating is 4 or 5 in our experiments). $n_{i}/n$ and $n_{Hi}/n_{i}$ represent the popularity and reputation of the product. The first part of (\ref{e1}) represents the probability of user $u$ knowing product $i$. For example, when $n_{u}$ or $n_{i}$ decreases, we assume the user is not familiar to this type of products or the product is less popular and thus the probability of user $u$ knowing product $i$ will decrease. The second part of (\ref{e1}) represents the probability of user $u$ being not interested in product $i$, which is determined by the popularity and reputation of the product. When user $u$ rates product $i$, $P_{ui}$ can be estimated from the rating score $R_{ui}$. For example, if $R_{ui}=1$, $P_{ui}=1$; if $R_{ui}=2$, $P_{ui}=0.8$; if $R_{ui}=3$, $P_{ui}=0.5$ and so on.

Given users $u$ and $v$ and the products which have not been rated by both of them, we can obtain the probability values as above. We use a matrix $\mathbf{A}^{\left(2\right)}$ to denote these probability values, and the second similarity measure between $u$ and $v$ can be calculated as follows:
\[
D_{uv}^{\left(2\right)}=\left|\sum_{z=1}^{m-\left|C_{uv}\right|}\left(A_{uz}^{\left(2\right)}-A_{vz}^{\left(2\right)}\right)\right|, \qquad S_{uv}^{\left(2\right)}=e^{-\frac{\gamma_{2}D_{uv}^{\left(2\right)}}{m-\left|C_{uv}\right|}} \; \left(\gamma_{2}>0\right)
\]
where $m-\left|C_{uv}\right|$ represents the size of the products having not been rated by both $u$ and $v$. Here we adopt a simple method to compute the difference between users $u$ and $v$, because the number of the products considered is usually large.
\subsubsection{Similarity Based on Rating Biases.}
We observe that users' rating values are usually unevenly distributed. For example, high ratings, 4 and 5, usually account for a large proportion, while low ratings, 1 and 2, hold a small proportion. We call this as the rating biases of users. Here, we adopt the idea of TF-IDF to measure the users' rating biases, and the user is viewed as the $document$ and the ratings are viewed as $words$. Matrix $\mathbf{A}^{\left(3\right)}$ is used to denote the rating biases and each element $A^{\left(3\right)}_{ur}$ is the user $u$'s bias for rating score $r\in \{1,2,3,4,5\}$. It can be calculated as follows:
\[
A_{ur}^{\left(3\right)} = rf\left(u,r\right)\times log_{base}\left(\frac{n}{uf\left(r\right)}\right), \qquad rf\left(u,r\right)=\frac{n_{ur}}{\begin{matrix}\sum_{z=1}^{5}n_{uz}\end{matrix}}
\]
where $uf\left(r\right)$ represents the number of users who have given the rating score $r$, $n_{ur}$ represents the frequency of rating score $r$ used in $u$'s rating history, and $base$ is a predefined parameter ($base=2$ in our experiments). One can see that $A^{\left(3\right)}_{ur}$ is proportional to $rf\left(u,r\right)$ and inversely proportional to $uf\left(r\right)$. We compute the third similarity measure as follows:
\[
D^{\left(3\right)}_{uv}=\left|\begin{matrix}\sum_{z=1}^{5}\left(A_{uz}^{\left(3\right)}-A_{vz}^{\left(3\right)}\right)\end{matrix}\right|, \qquad S_{uv}^{\left(3\right)}=e^{-\gamma_{3}D^{\left(3\right)}_{uv}} \; \left(\gamma_{3}>0\right)
\]

As above, for users $u$ and $v$, the three similarity measures $S_{uv}^{\left(1\right)},S_{uv}^{\left(2\right)},S_{uv}^{\left(3\right)} \in \left[0,1\right]$ and their weighted average $S_{uv}=\rho_{1}S_{uv}^{\left(1\right)}+\rho_{2}S_{uv}^{\left(2\right)}+\rho_{3}S_{uv}^{\left(3\right)}$ is used as the final rating behavior based user similarity, where $\rho_{1}$, $\rho_{2}$, $\rho_{3}$ are the weights to control the importance of the three parts.
\subsection{Rating Matrix Factorization}
We embed the user similarities in Sect. 4.1 into the matrix factorization model as a new regularization term. The insight is that two similar users should also be close to each other in the factorized latent feature space. We use the boldface uppercase letters $\mathbf{U}$ and $\mathbf{V}$ to denote the latent feature matrices of users and items, and our goal is solving the following minimization problem:
\begin{equation} \label{e2}
\begin{split}
&\min_{\mathbf{U},\mathbf{V}}\frac{1}{2}\sum_{u=1}^{n}\sum_{i=1}^{m}Y_{ui}\left(R_{ui}-\mathbf{U}_{u*}\mathbf{V}_{i*}^{T}\right)^{2}
+\frac{\alpha}{2}tr\left(\mathbf{U}\mathbf{U}^{T}\right)\\
&+\frac{\alpha}{2}tr\left(\mathbf{V}\mathbf{V}^{T}\right)+\frac{\beta}{2}\sum_{u=1}^{n}\sum_{v=u+1}^{n}S_{uv}\parallel\mathbf{U}_{u*}-\mathbf{U}_{v*}\parallel^{2}
\end{split}
\end{equation}
where $\mathbf{U}_{u*}$ and $\mathbf{V}_{v*}$ represent the latent features of user $u$ and product $i$, $\left(\cdot\right)^{T}$ and $tr\left(\cdot\right)$ denote the transposition and trace of a matrix, and $Y_{ui}$ is an indicator variable and its value is 1 if user $u$ rated product $i$ and 0 otherwise. $\alpha$ is the regularization parameter to prevent over-fitting while $\beta$ controlling the contribution from the user similarities. The insight of the last term in (\ref{e2}) is that if two users have very similar rating behaviors, it will put a larger penalty on the difference of the two users' latent features; otherwise it will put a smaller penalty on the difference of the latent features. To solve the problem conveniently, we transform the last term in (\ref{e2}) as follows,
\begin{flalign*}
&\quad \ \sum_{u=1}^{n}\sum_{v=u+1}^{n}S_{uv}\parallel\mathbf{U}_{u*}-\mathbf{U}_{v*}\parallel^{2} = \frac{1}{2}\sum_{u=1}^{n}\sum_{v=1}^{n}S_{uv}\parallel\mathbf{U}_{u*} - \mathbf{U}_{v*}\parallel^{2}\\
&= \frac{1}{2}\sum_{u=1}^{n}\sum_{v=1}^{n}S_{uv}\sum_{k=1}^{K}\left(U_{uk}-U_{vk}\right)^{2} = \sum_{k=1}^{K}\mathbf{U}_{*k}^{T}\mathbf{L}\mathbf{U}_{*k} = tr\left(\mathbf{U}^{T}\mathbf{L}\mathbf{U}\right)
\end{flalign*}
where $\mathbf{L}=\mathbf{D}-\mathbf{S}$ is the Laplacian matrix with $\mathbf{D}$ being a diagonal matrix whose diagonal element is $D_{uu}=\sum_{v=1}^{n}S_{uv}$. Thus, (\ref{e2}) becomes
\begin{equation} \label{e3}
\min_{\mathbf{U},\mathbf{V}}\frac{1}{2}\sum_{u=1}^{n}\sum_{i=1}^{m}Y_{ui}\left(R_{ui}-\mathbf{U}_{u*}\mathbf{V}_{i*}^{T}\right)^{2}+\frac{\alpha}{2}tr\left(\mathbf{V}\mathbf{V}^{T}\right)
+\frac{1}{2}tr\left[\mathbf{U}^{T}\left(\alpha\mathbf{I}+\beta\mathbf{L}\right)\mathbf{U}\right]
\end{equation}
where $\mathbf{I}$ is an identity matrix. We apply the alternating gradient descent to optimize one column of $\mathbf{U}$ or one row of $\mathbf{V}$ at a time. If we use $\mathrm{F}$ to represent the objective function in (\ref{e3}), the gradients can be computed as follows:
\[
\frac{\partial \mathrm{F}}{\partial \mathbf{U}_{*k}} = \left(\alpha\mathbf{I}+\beta\mathbf{L}\right)\mathbf{U}_{*k}-\mathbf{x} \qquad \qquad \qquad \qquad  \quad \ \,
\]
\[
\frac{\partial \mathrm{F}}{\partial \mathbf{V}_{i*}} =
-\begin{matrix}\sum_{u=1}^{n}Y_{ui}\left(R_{ui}-\mathbf{U}_{u*}\mathbf{V}_{i*}^{T}\right)\mathbf{U}_{u*}\end{matrix}+\alpha\mathbf{V}_{i*}
\]
where $\mathbf{x}$ is a $n \times 1$ vector whose element is $x_{u}=\begin{matrix}\sum_{i=1}^{m}Y_{ui}\left(R_{ui}-\mathbf{U}_{u*}\mathbf{V}_{i*}^{T}\right)V_{ik}\end{matrix}$.
\section{Neighborhood Based Latent Feature Mapping}
 The proposed MFUS can learn the domain-specific latent features of users in different domains. However, for the cold-start users $U_{T}$, we can only obtain their latent features in the auxiliary domain which cannot be used directly for making recommendation in the target domain due to the different semantic meanings of latent features in different domains. However, the same user's latent features in different domains can be highly correlated. For example, if a user likes martial arts novels, he/she may also be interested in Chinese swordsman films. Therefore, we try to use the linked users $U_{L}$ as a bridge to learn the function $\mathcal{F}$ which can map the user's latent features from the auxiliary domain to the target domain. The input of the mapping function $\mathcal{F}$ is a user's latent features in the auxiliary domain and the output is the same user's latent features in the target domain.

We adopt the Gradient Boosting Trees (GBT) method \cite{GBT} to learn the mapping function $\mathcal{F}$ since it is powerful to capture higher-order transformation relationship between the input and output. GBT is a function approximation method which applies numerical optimization in function space rather than the parameter space. Given a training set $\left\{\bm{x}^{i}, y^{i}\right\}_{i=1}^{N}$ where $\bm{x}^{i}\in R^{K\times1}$ and $y^{i}\in R$, the goal of GBT is to find a function $f\left(\bm{x}\right)$ which makes the expected value of a specified loss function $\Psi\left(f\left(\bm{x}\right)\right)$ minimized over the training set and can predict a response value $y\in R$ for a new $\bm{x}\in R^{K\times1}$. Specifically, the finally returned $f\left(\bm{x}\right)$ is built in a stagewise process by performing gradient descent in the function space. At the $m$th boosting,
\begin{equation} \label{e4}
f_{m}\left(\bm{x}\right)=f_{m-1}\left(\bm{x}\right)+\nu \eta_{m}h_{m}\left(\bm{x}; \boldsymbol{\alpha}_{m}\right)
\end{equation}
where $h_{m}\left(\bm{x}; \boldsymbol{\alpha}_{m}\right)$ is a function parameterised by $\boldsymbol{\alpha}_{m}$, $\eta_{m}$ is the learning rate, and $0<\nu\leq1$ is the shrinkage parameter to prevent over-fitting. The learning procedure consists of two alternative steps in the $m$th iteration: first fit a new component function $h_{m}\left(\bm{x}; \boldsymbol{\alpha}_{m}\right)$ according to the ``pseudo-response'' remained in the $\left(m-1\right)$th iteration and then the ``line search'' is performed to derive $\eta_{m}$. In our experiments, we use the squared error function and set $\nu=0.01$, $\eta_{m}=1$.

Assuming the dimension of the latent features in the target domain is $K_{t}$, we can use GBT $K_{t}$ times and learn the mapping function $\mathcal{F}=\left\{f^{\left(k\right)}\left(\bm{x}\right)\right\}^{K_{t}}_{k=1}$,
where the $j$th subfunction $f^{\left(j\right)}\left(\bm{x}\right)$ takes the user's latent features in the auxiliary domain as input and returns the $j$th mapped latent feature in the target domain. Considering that the users with similar rating behaviors should share similar latent features, for each cold-start user, we use the similar linked users to learn the mapping function. Thus in the last step of our model, for each user $u\in U_{T}$, we use $\mathbb{N}_{u}$ to denote the similar linked users to $u$ with each $v\in\mathbb{N}_{u}$, $S^{a}_{uv}>sim$. Here, $sim$ is a predefined similarity threshold value and $S^{a}_{uv}$ is the user similarity in the auxiliary domain computed in MFUS. Latent feature pairs $\left\{\mathbf{U}^{a}_{v*},\mathbf{U}^{t}_{v*}\right\}_{v\in \mathbb{N}_{u}}$, where $\mathbf{U}^{a}_{v*}$ and $\mathbf{U}^{t}_{v*}$ represent user $v$'s latent features in the auxiliary domain and target domain, are used to learn the mapping function $\mathcal{F}_{u}=\left\{f^{\left(k\right)}_{u}\left(\bm{x}\right)\right\}^{K_{t}}_{k=1}$ via GBT. According to the latent features $\mathbf{U}^{a}_{u*}$ and the mapping function $\mathcal{F}_{u}$, we can compute the user mapped latent features $\emph{\textbf{u}}$ in the target domain with the element $u_{k}=f^{\left(k\right)}_{u}\left(\mathbf{U}^{a}_{u*}\right)$. Based on the mapped latent features $\emph{\textbf{u}}$ and the latent features of items in the target domain $\mathbf{V}^{t}$, we can get the rating predictions by
\begin{equation} \label{e5}
\hat{\mathbf{r}}=\mathbf{V}^{t}\emph{\textbf{u}}^{T}
\end{equation}

The pseudocode of the proposed CDLFM model is given in Algorithm 1.
\begin{algorithm} [t]
\caption{\small CDLFM: $\mathbf{C}$ross-$\mathbf{D}$omain $\mathbf{L}$atent $\mathbf{F}$eature $\mathbf{M}$apping}
\hspace*{0.02in}{\bf \small Input:\ \ \ }
\small
$\mathbf{R}^{t}$, $\mathbf{R}^{a}$ - rating matrices in the target domain and auxiliary domain\\
\hspace*{0.02in}{\bf \quad \quad \quad \  }
\small
$K_{t}$, $K_{a}$ - feature dimensions in the target domain and auxiliary domain\\
\hspace*{0.02in}{\bf \quad \quad \quad \  }
\small
$sim$ - the similarity threshold value for picking nearest linked users\\
\hspace*{0.02in}{\bf \small Output:}
\small
$\hat{\mathbf{R}}$ - matrix of rating prediction whose rows are $U_{T}$ and columns are $P_{t}$
\begin{algorithmic}[1]
\small
\State obtain the latent feature matrices $U^{t}$, $V^{t}$ in the target domain via MFUS
\State obtain the user similarity matrix $\mathbf{S}^{a}$ and the user latent feature matrix $U^{a}$ in the auxiliary domain via MFUS
\For{every cold-start user $u$ in $U_{T}$}
\State Find his/her nearest linked users $\mathbb{N}_{u}$ according to $\mathbf{S}^{a}_{u*}$ and $sim$
\For{each dimension $k$ of the user latent feature vector in the target domain}
\State Construct the training set $\mathbf{T}_{u}^{\left(k\right)}=\left\{\mathbf{U}^{a}_{v*},\mathbf{U}^{t}_{vk}\right\}_{v\in \mathbb{N}_{u}}$
\State Initialize $f_{u}^{\left(k\right)}\left(\bm{x}\right)=f_{u0}^{\left(k\right)}\left(\bm{x}\right)$
\While{the objective loss function $\Psi\left(f^{\left(k\right)}_{u}\left(\bm{x}\right)\right)$ has not been convergent}
\State Compute the current ``pseudo-response''
\State Learn a new function $h\left(\bm{x}; \boldsymbol{\alpha}\right)$ to fit the ``pseudo-response''
\State Update $f^{\left(k\right)}_{u}\left(\bm{x}\right)$ with (\ref{e4})
\EndWhile
\EndFor
\State Predict $u$'s mapped latent features with $\mathcal{F}_{u}=\left\{f^{\left(k\right)}_{u}\left(\bm{x}\right)\right\}^{K_{t}}_{k=1}$ and $\mathbf{U}^{a}_{u*}$
\State  Compute the row in $\hat{R}$ about $u$ with (\ref{e5})
\EndFor
\State \Return $\hat{\mathbf{R}}$
\end{algorithmic}
\end{algorithm}
\section{Experiments}
\subsection{Experiment Setup}
We extract two datasets from the Amazon rating data \cite{Ups and downs} in which multiple item domains are contained. The first extracted dataset consists of the ratings in the movie domain and the book domain, and the second one consists of the ratings about movies and electronic products. We first filter out the linked users and items with very small number of ratings. In order to gain better experiment performance, besides the linked users, some active users who have given a large number of ratings in a certain domain are also included. Finally, in the first dataset, we have 16926 linked users, 1000 movie active users and 500 book active users. In the second dataset, we have 12004 linked users, 199 movie active users and 724 electronics active users. The statistics of the two datasets are given in Table \ref{t1}. And We compare our model CDLFM with the following baselines:
\begin{table}
\footnotesize
\centering
\caption{Statistics of the two datasets used for evaluation}
\begin{tabular}{|p{2.2cm}|p{2.2cm}|p{2cm}|p{2cm}|p{2cm}|}
\hline
Dataset 1 & Rating value & \multicolumn{2}{c|}{} & Density\\
\hline
\multirow{3}{1.2cm}{Movie} & \multirow{3}{*}{$\left\{1,2,3,4,5\right\}$} & $\, \#$users & $\,$17926 & \multirow{3}{*}{0.00225} \\
\cline{3-4}
& &$\, \#$movies & $\,$4595 & \\
\cline{3-4}
& & $\, \#$ratings & $\,$185421 & \\
\hline
\multirow{3}{*}{Book} & \multirow{3}{*}{$\left\{1,2,3,4,5\right\}$} & $\, \#$users & $\,$17426 & \multirow{3}{*}{0.00149} \\
\cline{3-4}
& & $\, \#$books & $\,$8935 &  \\
\cline{3-4}
& & $\, \#$ratings & $\,$231564 & \\
\hline
Dataset 2 & \multicolumn{4}{c|}{} \\  
\hline
\multirow{3}{*}{Movie} & \multirow{3}{*}{$\left\{1,2,3,4,5\right\}$} & $\, \#$users & $\,$12203 & \multirow{3}{*}{0.00307} \\
\cline{3-4}
& & $\, \#$movies & $\,$3625 & \\
\cline{3-4}
& & $\, \#$ratings & $\,$135587 & \\
\hline
\multirow{3}{*}{Electronics} & \multirow{3}{*}{$\left\{1,2,3,4,5\right\}$} & $\, \#$users & $\,$12728 & \multirow{3}{*}{0.00212} \\
\cline{3-4}
& & $\, \#$electronics & $\,$4302 & \\
\cline{3-4}
& & $\, \#$ratings & $\,$115955 & \\
\hline
\end{tabular}
\label{t1}
\end{table}
\begin{itemize}
\item[$\bullet$] AF: Average Filling is a heuristic method used in \cite{Transfer in CF}, which estimates with the sum of global average rating, user bias and item bias.
\item[$\bullet$] CDCF-U: It is a user-based neighborhood Cross-Domain Collaborative Filtering model used in \cite{CDTF}.
\item[$\bullet$] CDCF-I: It is an item-based neighborhood Cross-Domain Collaborative Filtering model used in \cite{CDTF}.
\item[$\bullet$] CMF \cite{CMF}: CMF is a transfer learning method in which the user latent features are shared between different domains.
\item[$\bullet$] TMatrix \cite{TMatrix}: It achieves the features mapping across domains via a learned Transformation Matrix based on linked users. In our experiments, the latent features are learned via MF \cite{Probabilistic}.
\item[$\bullet$] EMCDR \cite{EMCDR}: It is one state-of-the-art cross-domain recommendation method for cold-start users. In EMCDR, latent features are learned by MF firstly, and then MLP is used for latent space mapping.
\end{itemize}
We do not make comparison with CDTF \cite{CDTF} due to the serious sparsity of our datasets, which degrades the effectiveness of CDTF. Root Mean Square Error (RMSE) and Mean Absolute Error (MAE) \cite{Transfer in CF} are used as the evaluation metrics.
\subsection{Experimental Results}
Experiments with different auxiliary and target domains are denoted as BM, MB, EM and ME for brevity. For example, BM denotes the experiments on Dataset 1 with $\mathbf{B}$ooks as the auxiliary domain and $\mathbf{M}$ovies as the target domain. The dimension of latent features is set to 15 and $sim$ in CDLFM is set to 0.45.
\subsubsection{Impact of Data Density.}
Firstly, we evaluate these methods under different data density levels. We randomly select 50$\%$ of the total linked users as the cold-start users whose ratings in the target domain compose the test set and the remaining linked users are in the training set. To simulate different density levels, we construct three different training sets denoted as density levels 50$\%$, 70$\%$ and 100$\%$. Taking the density level 70$\%$ for example, the training set consists of 70$\%$ of the total ratings in the auxiliary domain and 70$\%$ of the remaining ratings (after removing the cold-start users' ratings) in the target domain.

\begin{figure}[t]
\subfigure[BM]{\includegraphics[height=2.99cm, width=2.99cm, angle=0]{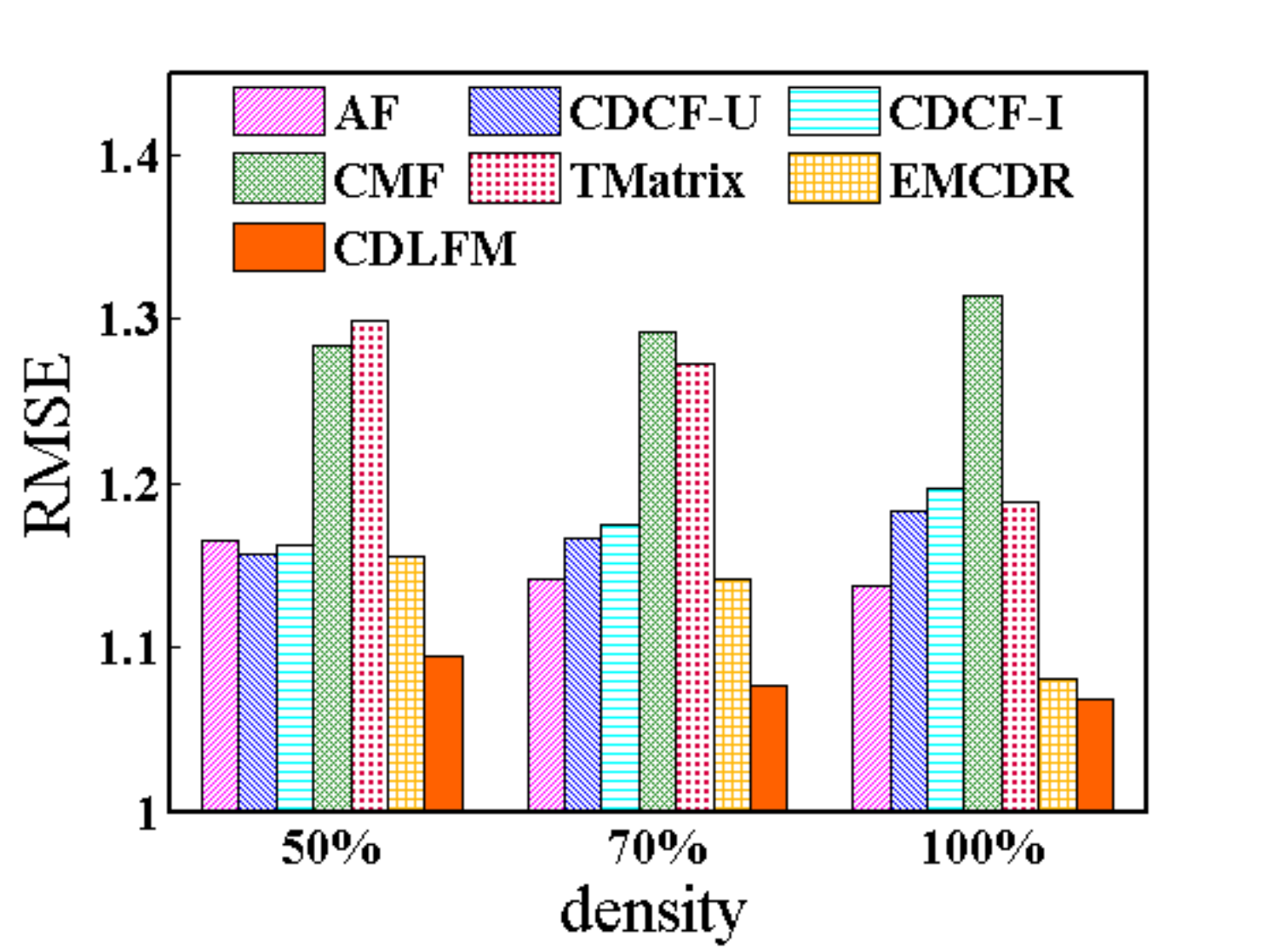}}
\subfigure[MB]{\includegraphics[height=2.99cm, width=2.99cm, angle=0]{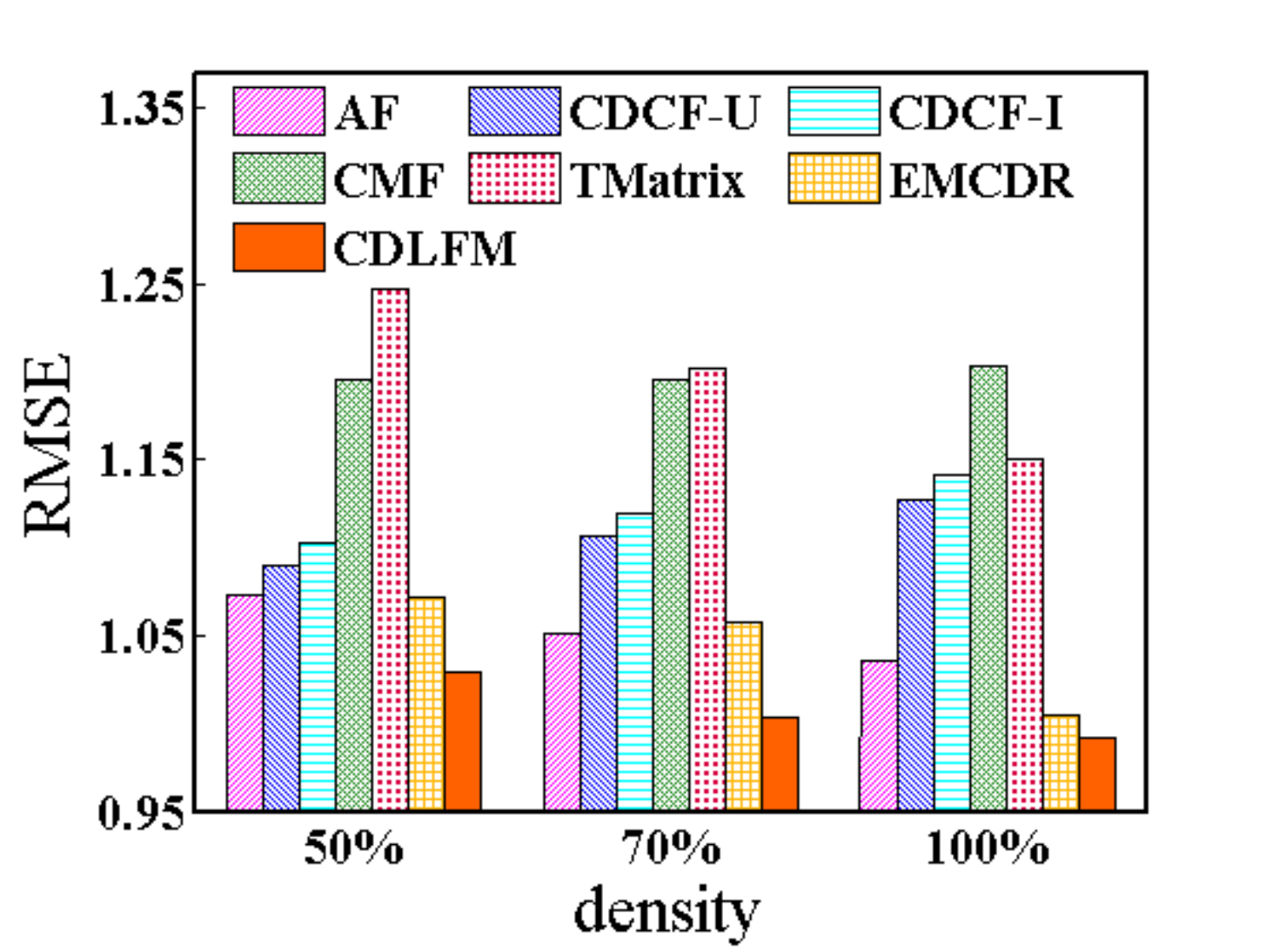}}
\subfigure[EM]{\includegraphics[height=2.99cm, width=2.99cm, angle=0]{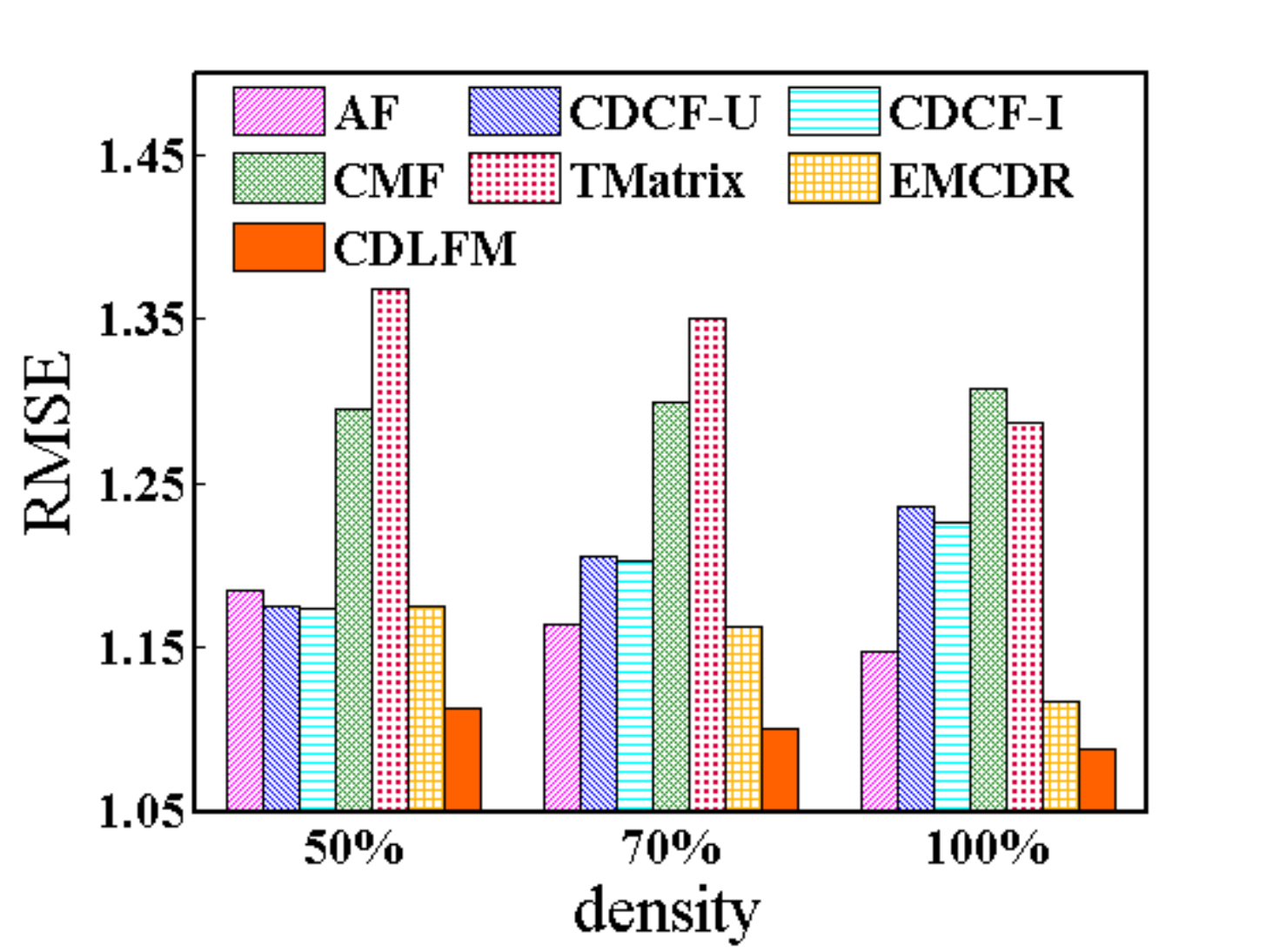}}
\subfigure[ME]{\includegraphics[height=2.99cm, width=2.99cm, angle=0]{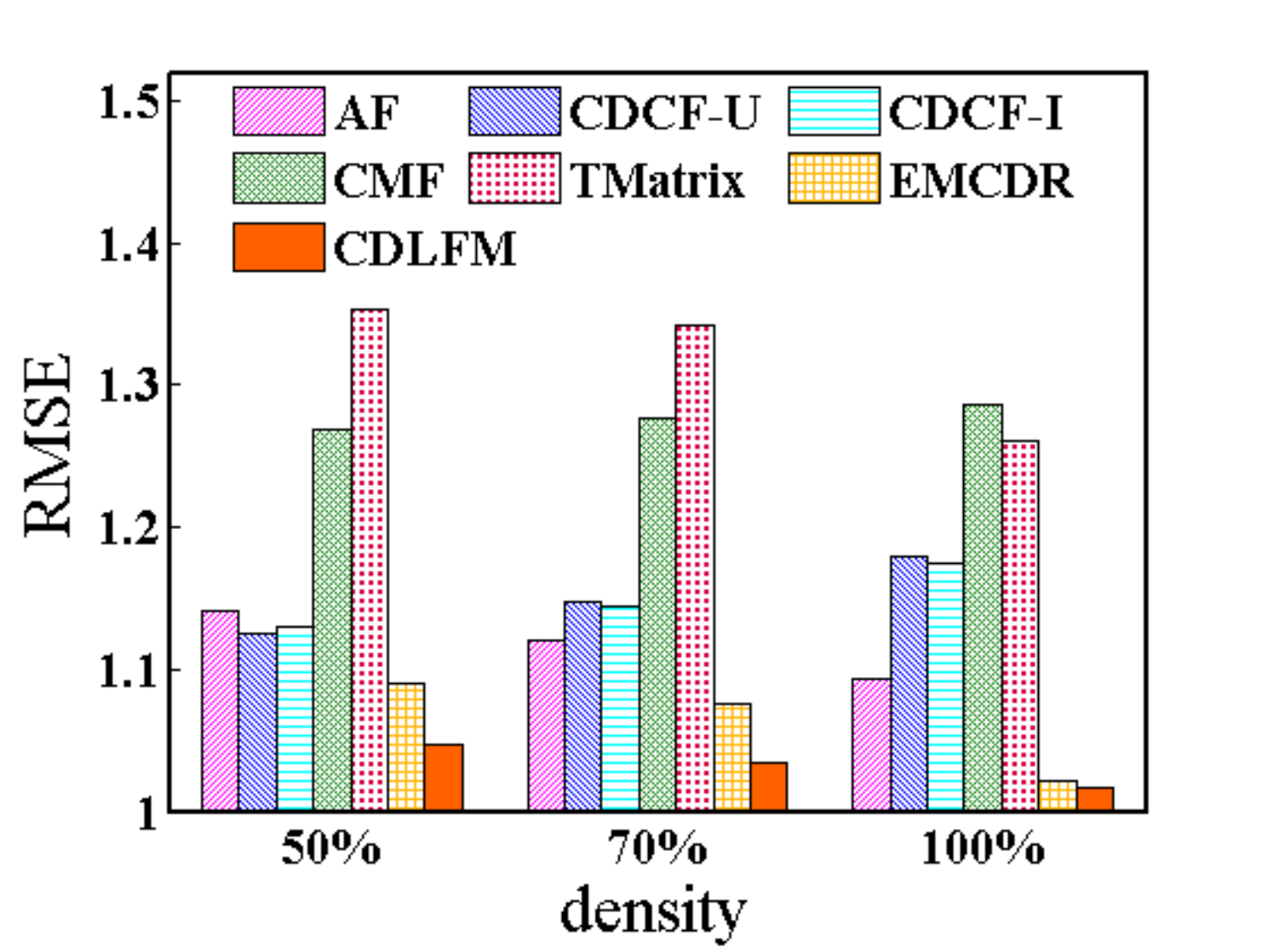}}
\caption{RMSE of methods under different data density levels}
\label{f3}
\end{figure}

\begin{figure}[t]
\subfigure[BM]{\includegraphics[height=2.99cm, width=2.99cm, angle=0]{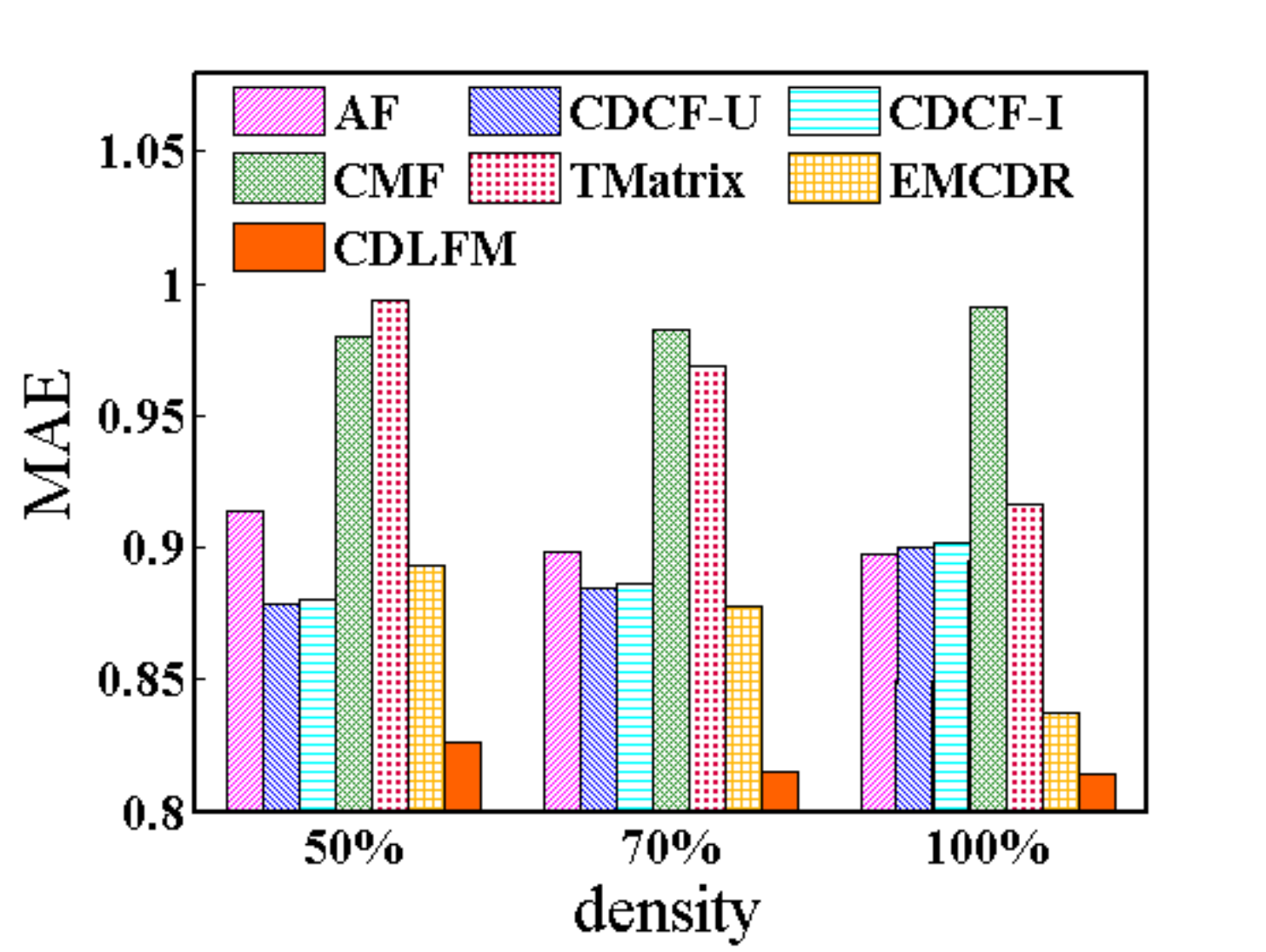}}
\subfigure[MB]{\includegraphics[height=2.99cm, width=2.99cm, angle=0]{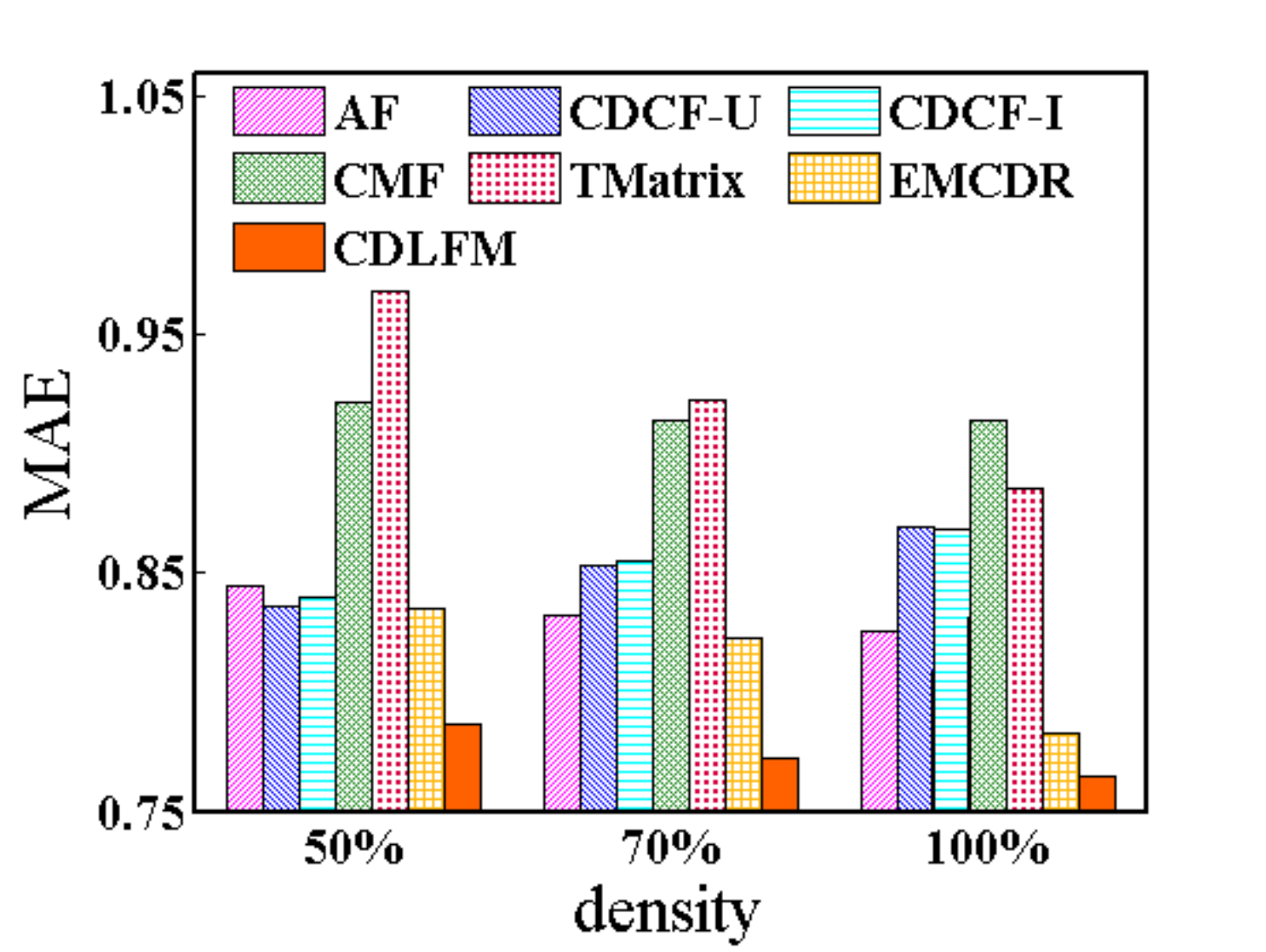}}
\subfigure[EM]{\includegraphics[height=2.99cm, width=2.99cm, angle=0]{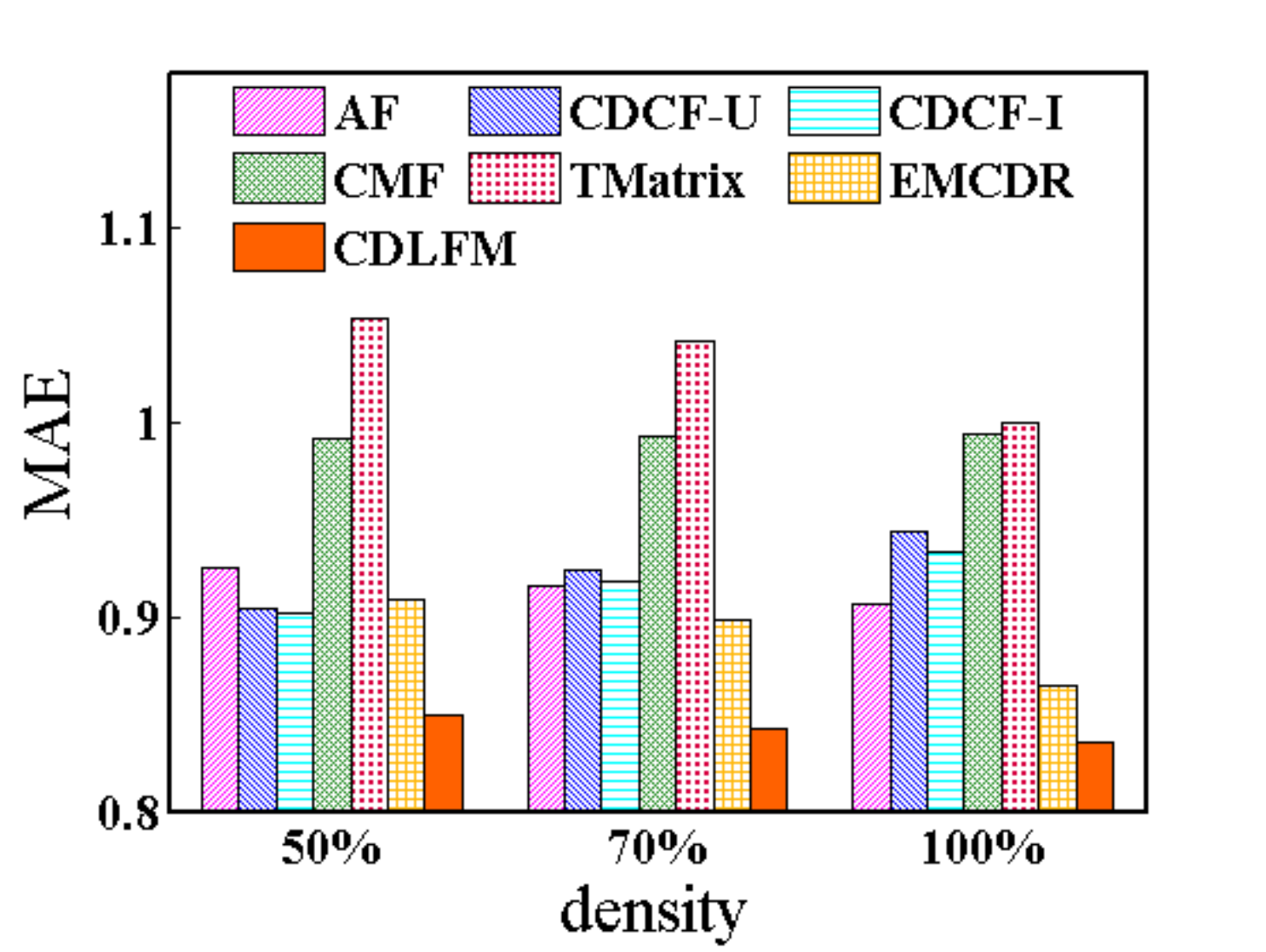}}
\subfigure[ME]{\includegraphics[height=2.99cm, width=2.99cm, angle=0]{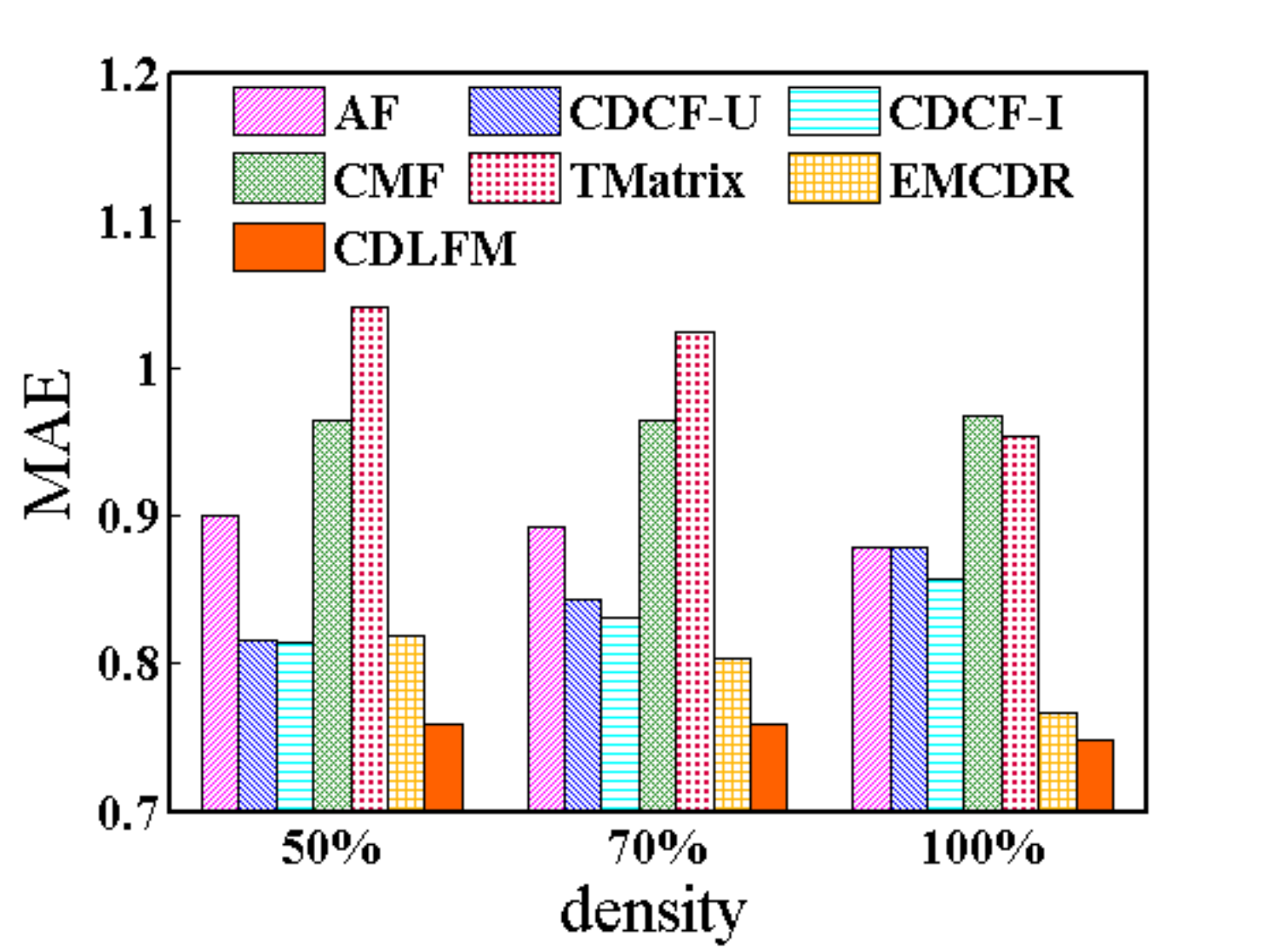}}
\caption{MAE of methods under different data density levels}
\label{f4}
\end{figure}

Figures \ref{f3} and \ref{f4} report the results on different datasets and different data density levels. One can see that our CDLFM model performs best under all different data density levels. For AF and neighborhood based methods, they cannot capture the global characteristics of users and items as factorization models do. Besides, user's preference and rating behaviors are varied in different domains, and they cannot get better performance because of not considering domain-specifically, and so does CMF. In TMatrix, the transformation matrix is a linear mapping function which is not capable to model the non-linear relationship between different domains' latent features. For EMCDR, MLP are learned based on all linked users which may introduce noise. Besides, from Fig. \ref{f3} and \ref{f4}, we can also see that the sparser the dataset is, the improvement of our model compared to EMCDR is more obvious.

In our model, MFUS takes users' rating behaviors into consideration which can alleviate the data sparsity and learn more accurate domain-specific latent features. Then neighborhood based GBT learns the user-specific higher-order feature mapping function via similar linked users. Therefore, we can predict cold-start users' latent features and preference accurately in the target domain.
\subsubsection{Impact of the Size of Linked Users.}
To evaluate the impact of the size of linked users, we conduct our experiments with three different user overlap levels between the two domains, namely 30$\%$, 50$\%$ and 70$\%$. Taking overlap level 30$\%$ for example, we randomly select 70$\%$ of the total linked users as the cold-start users, and the remaining ratings (after removing the cold-start users' ratings in the target domain) in the dataset compose the training set.

\begin{figure}[t]
\subfigure[BM]{\includegraphics[height=2.99cm, width=2.99cm, angle=0]{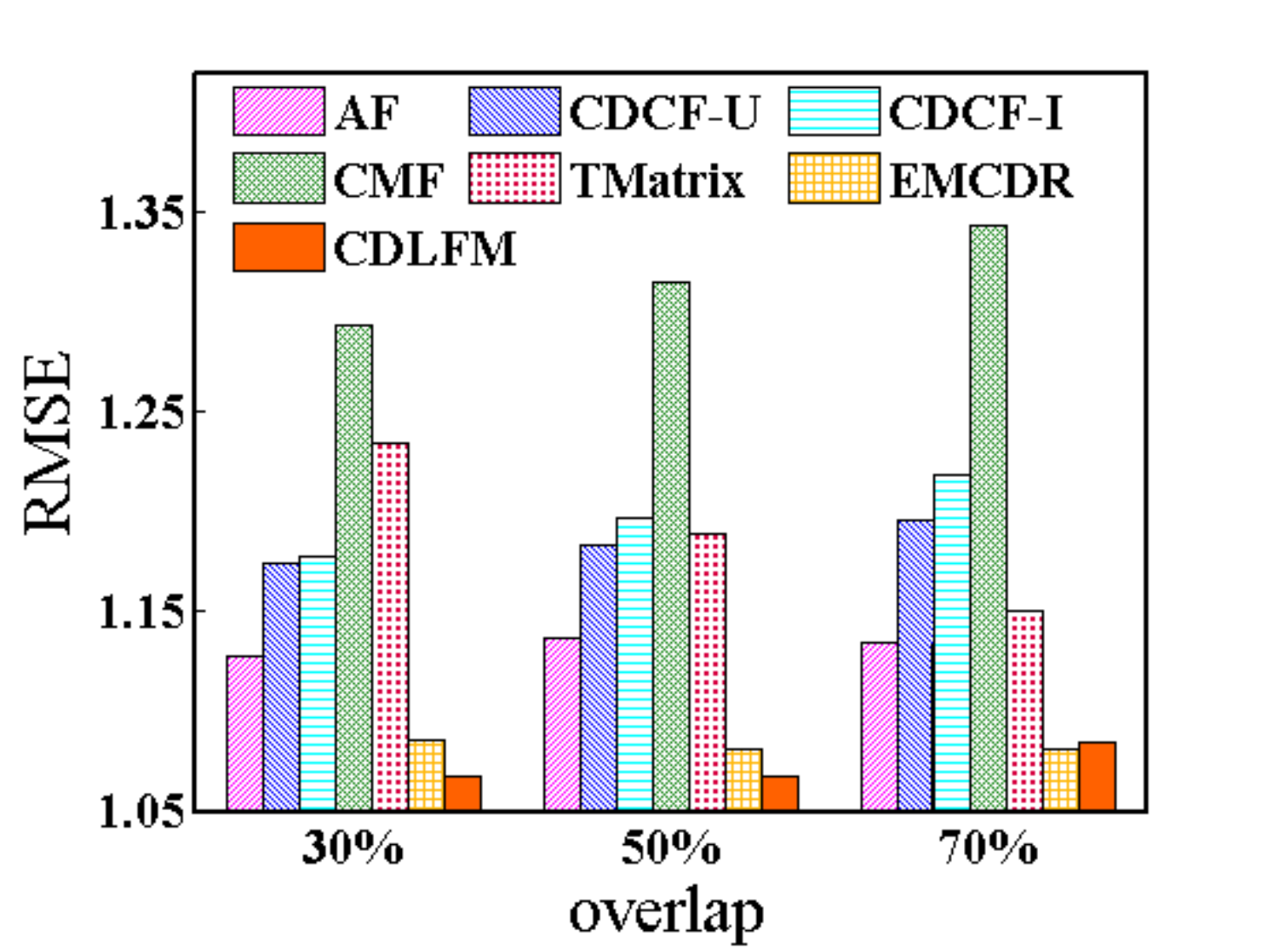}}
\subfigure[MB]{\includegraphics[height=2.99cm, width=2.99cm, angle=0]{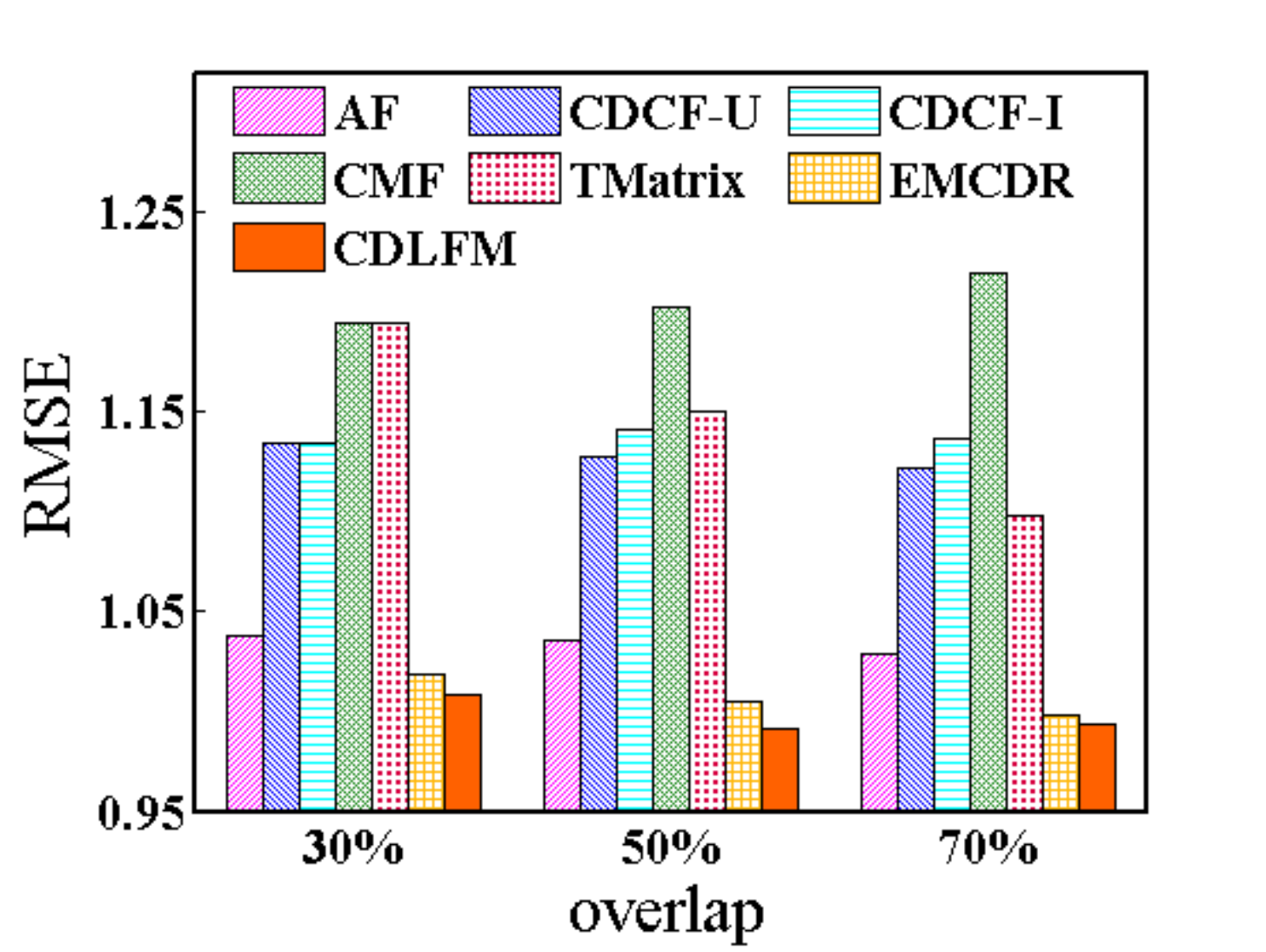}}
\subfigure[EM]{\includegraphics[height=2.99cm, width=2.99cm, angle=0]{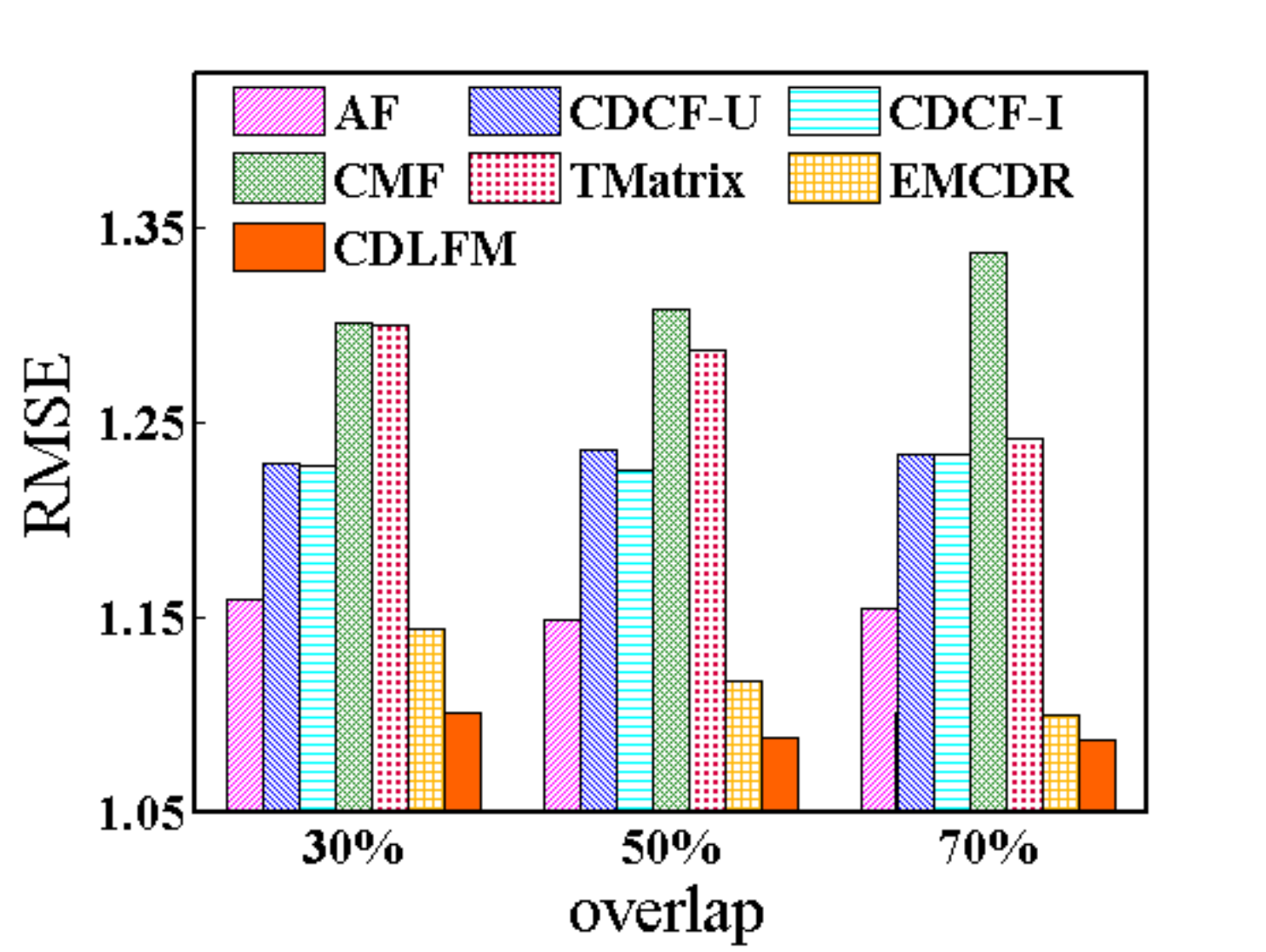}}
\subfigure[ME]{\includegraphics[height=2.99cm, width=2.99cm, angle=0]{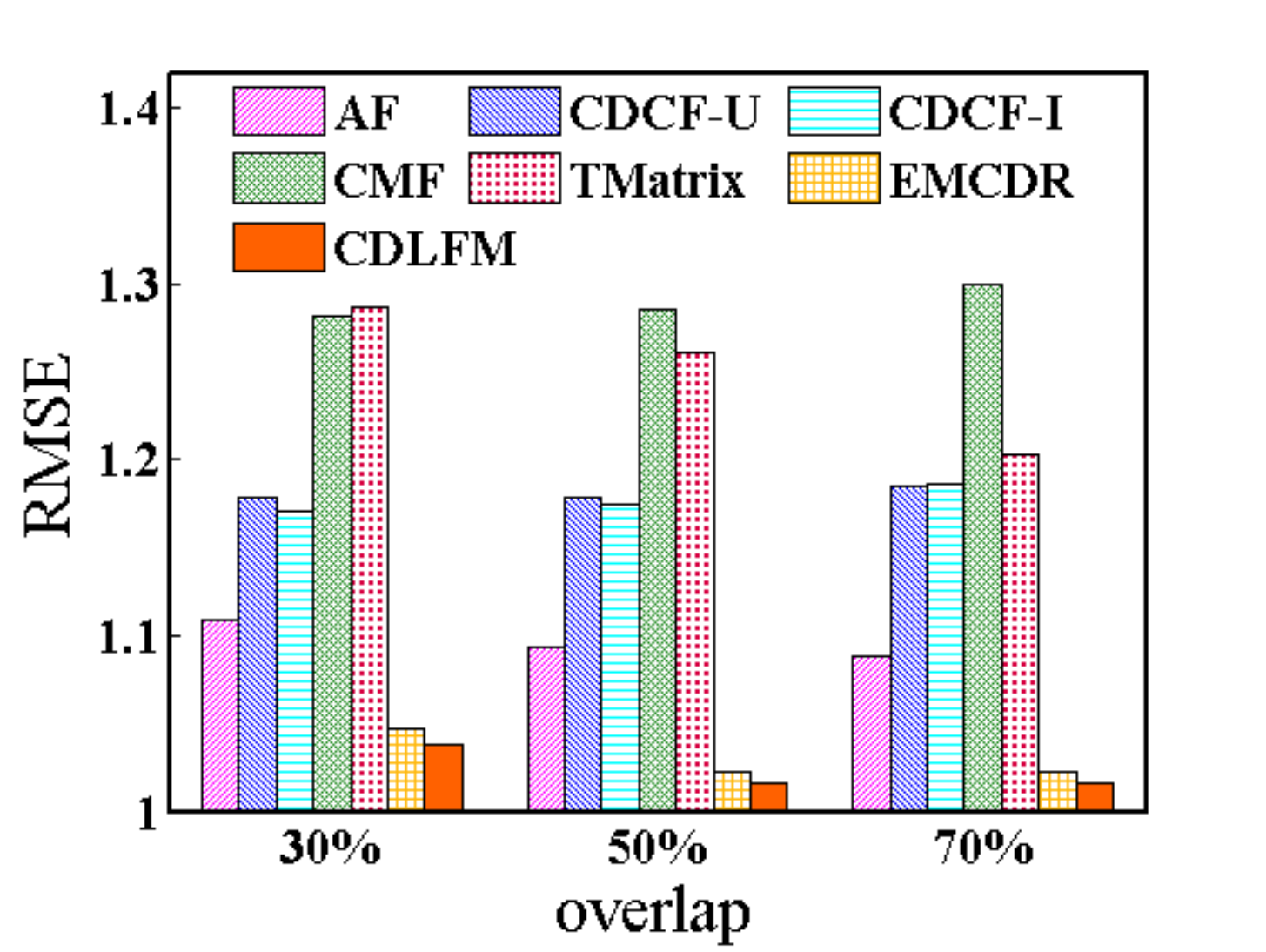}}
\caption{RMSE of methods under different overlap levels}
\label{f5}
\end{figure}

\begin{figure}[t]
\subfigure[BM]{\includegraphics[height=2.99cm, width=2.99cm, angle=0]{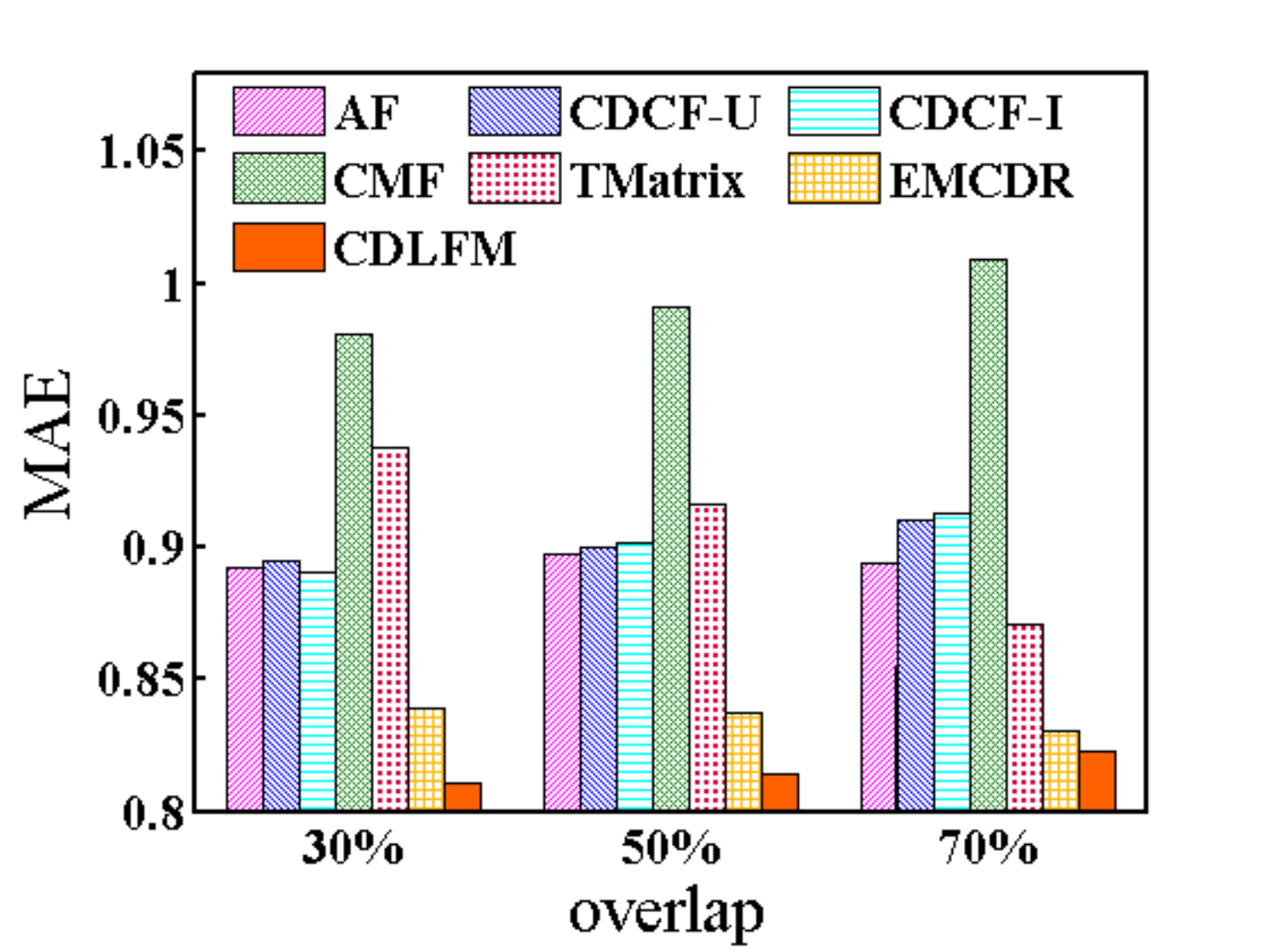}}
\subfigure[MB]{\includegraphics[height=2.99cm, width=2.99cm, angle=0]{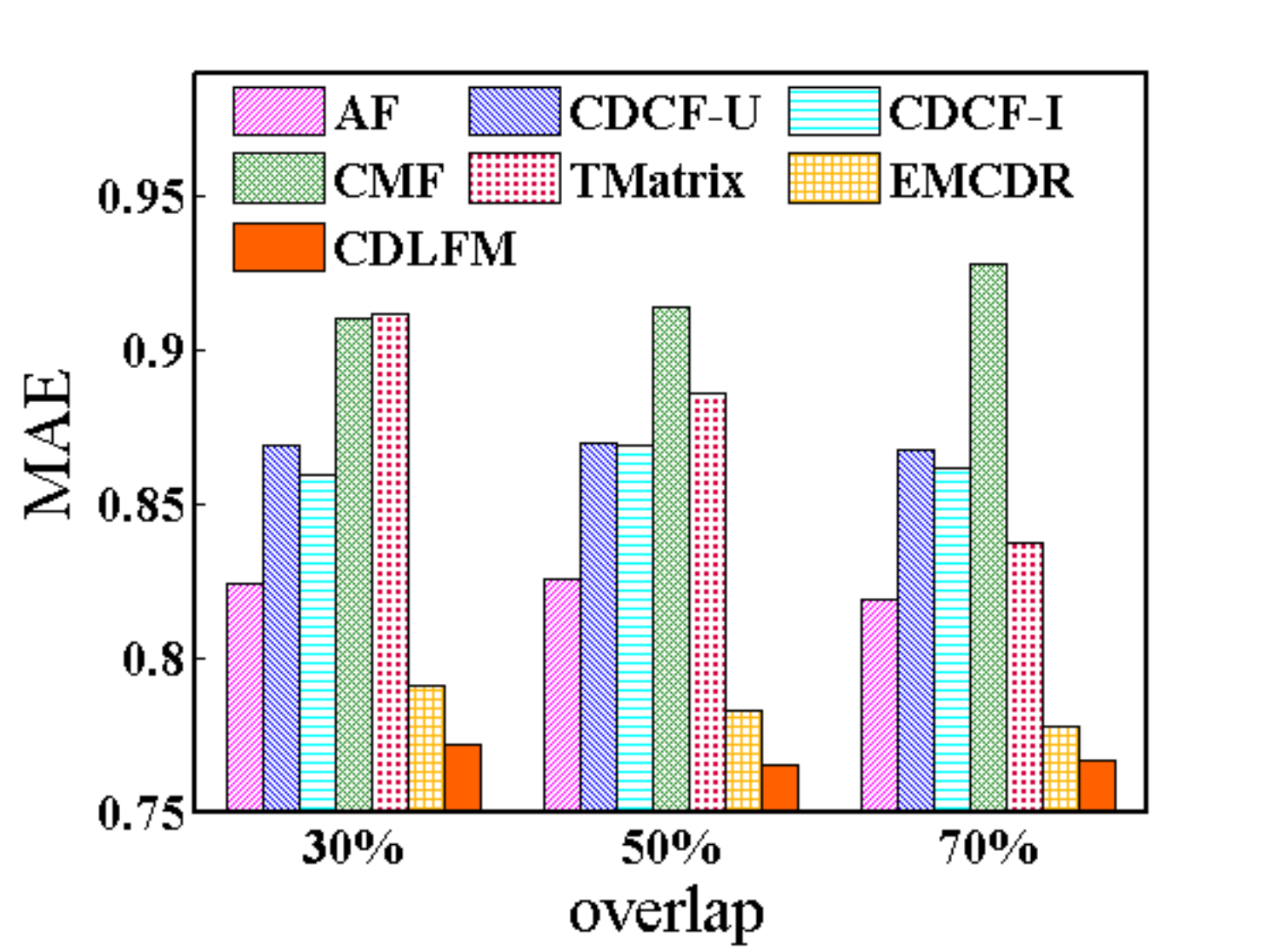}}
\subfigure[EM]{\includegraphics[height=2.99cm, width=2.99cm, angle=0]{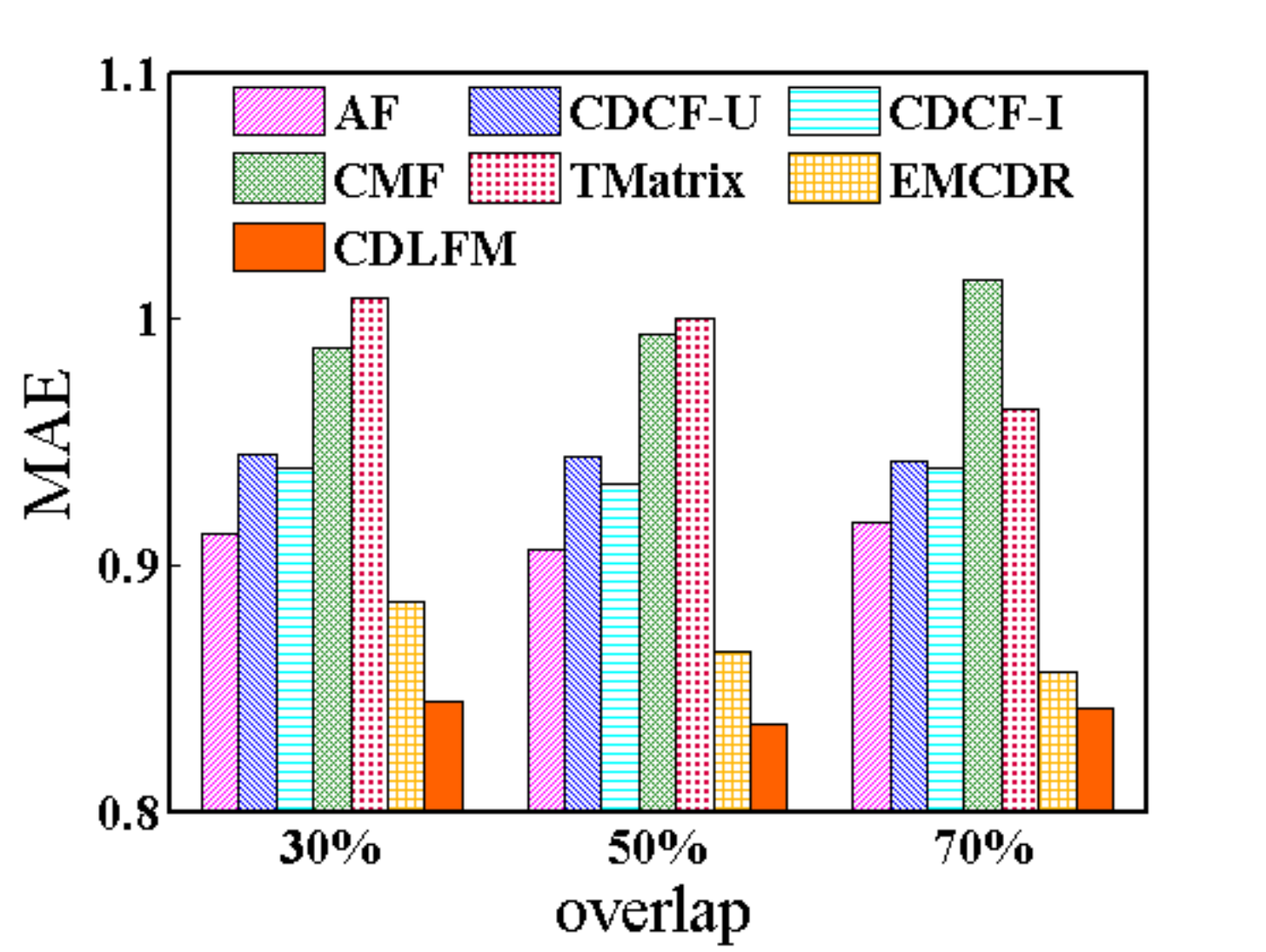}}
\subfigure[ME]{\includegraphics[height=2.99cm, width=2.99cm, angle=0]{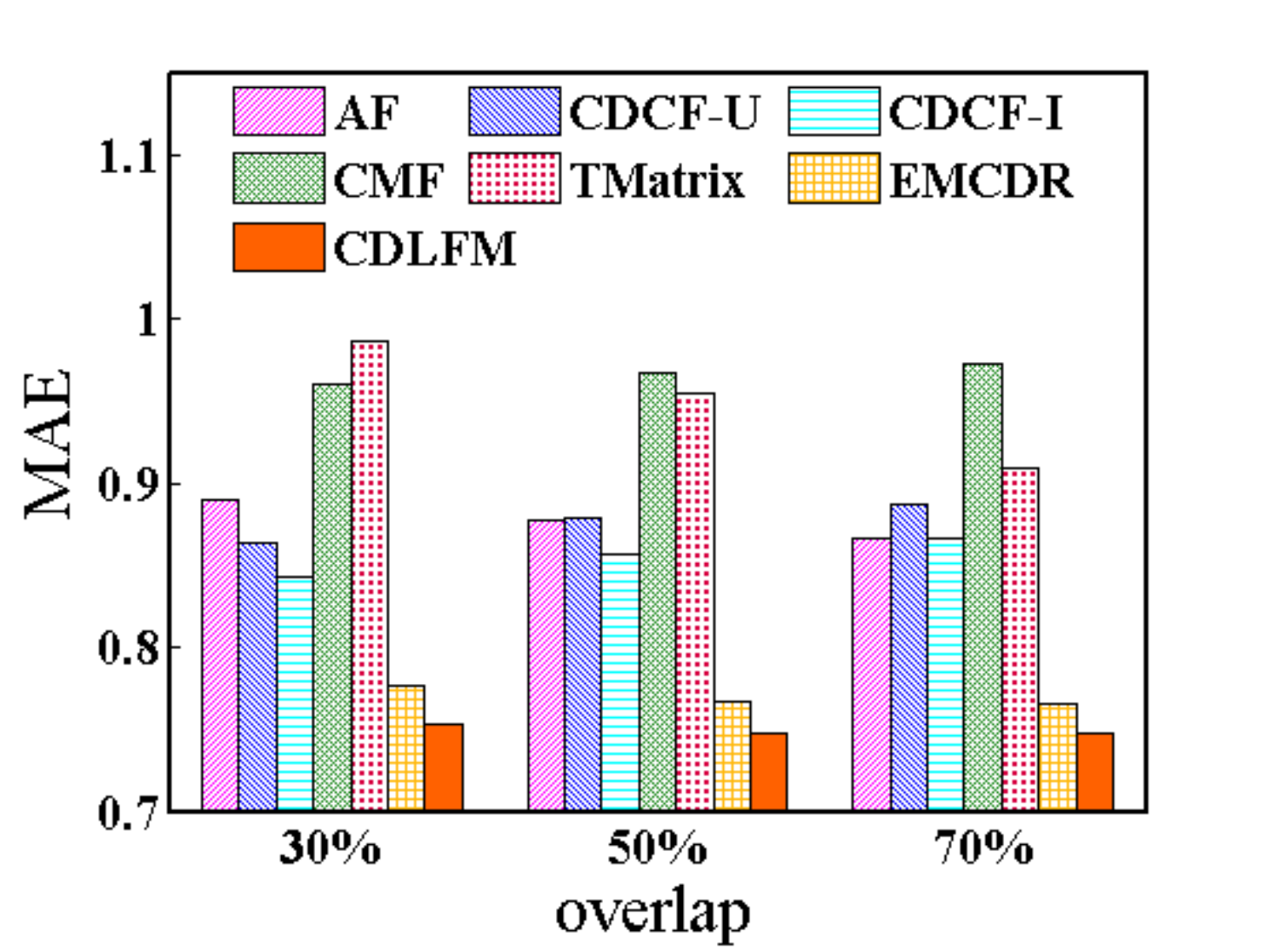}}
\caption{MAE of methods under different overlap levels}
\label{f6}
\end{figure}

 The results are reported in Fig. \ref{f5} and \ref{f6}. One can see that our model achieves the best performance under all user overlap levels. Similarly, the less users overlap between two domains, the improvement of our model compared to EMCDR is more obvious, which means our CDLFM model is good at capturing valuable knowledge from small amount of data.
\subsubsection{Comparison with Two Variants.}
\begin{figure}[b]
\subfigure[BM]{\includegraphics[height=2.99cm, width=2.99cm, angle=0]{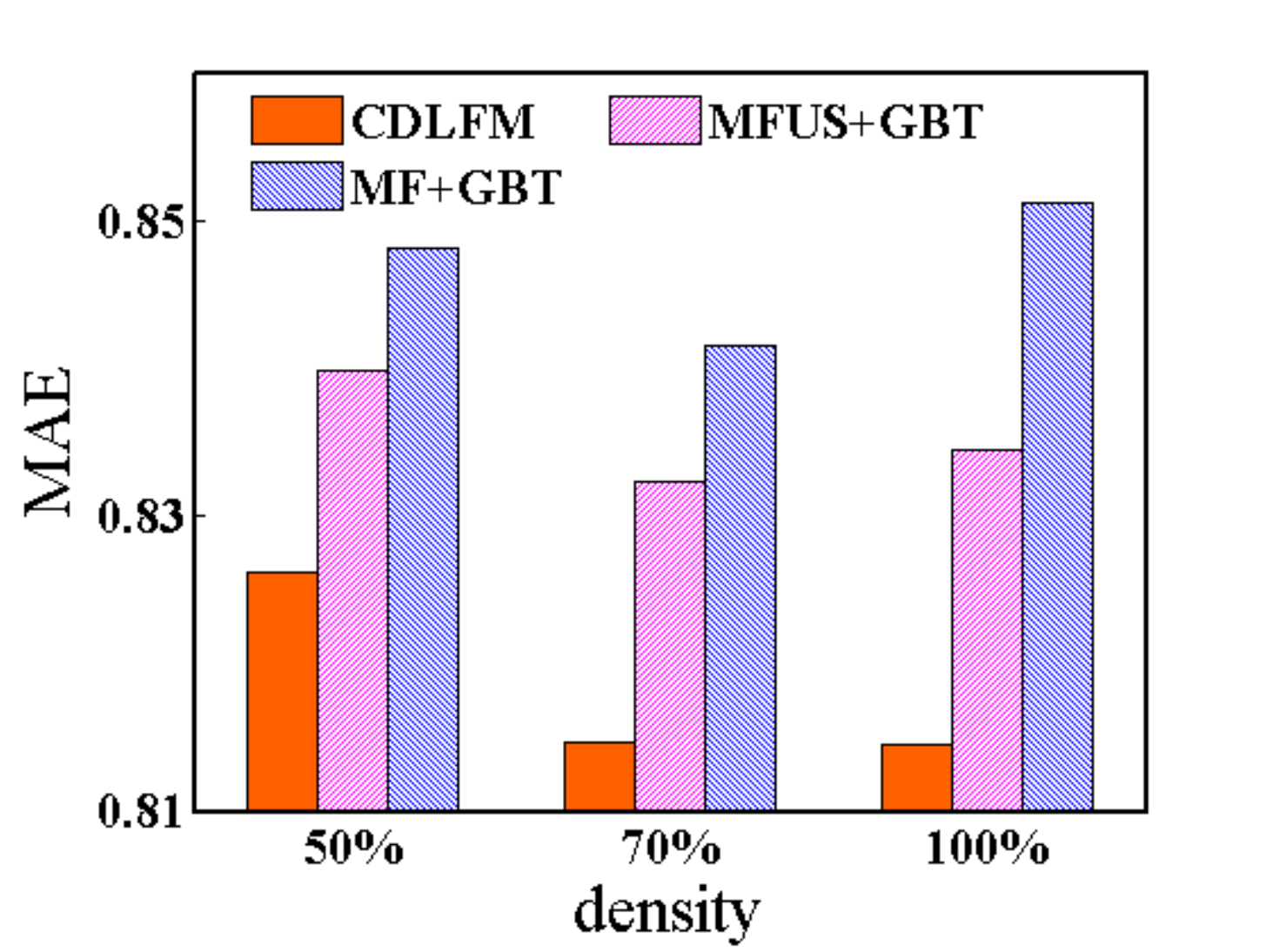}}
\subfigure[MB]{\includegraphics[height=2.99cm, width=2.99cm, angle=0]{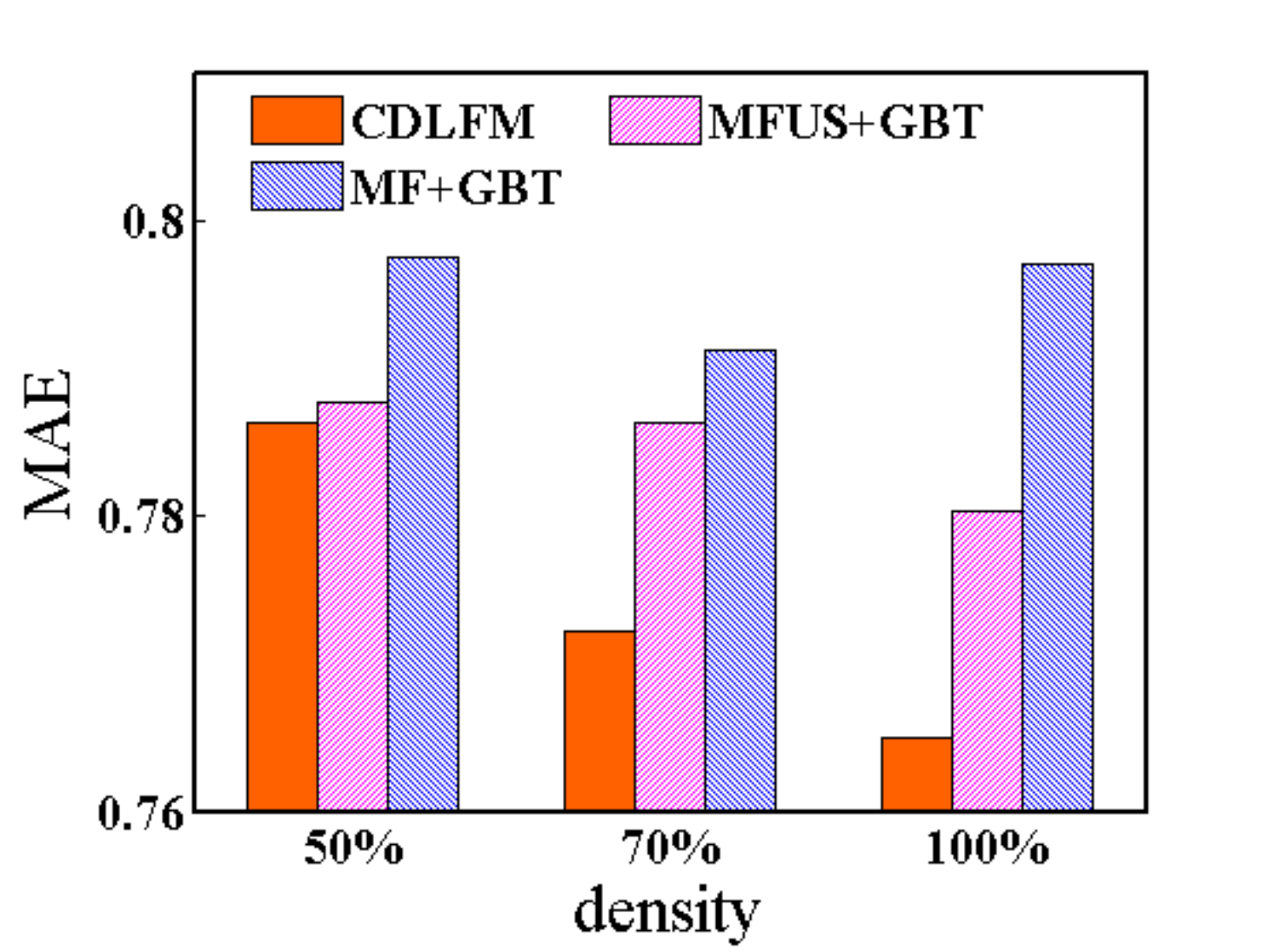}}
\subfigure[BM]{\includegraphics[height=2.99cm, width=2.99cm, angle=0]{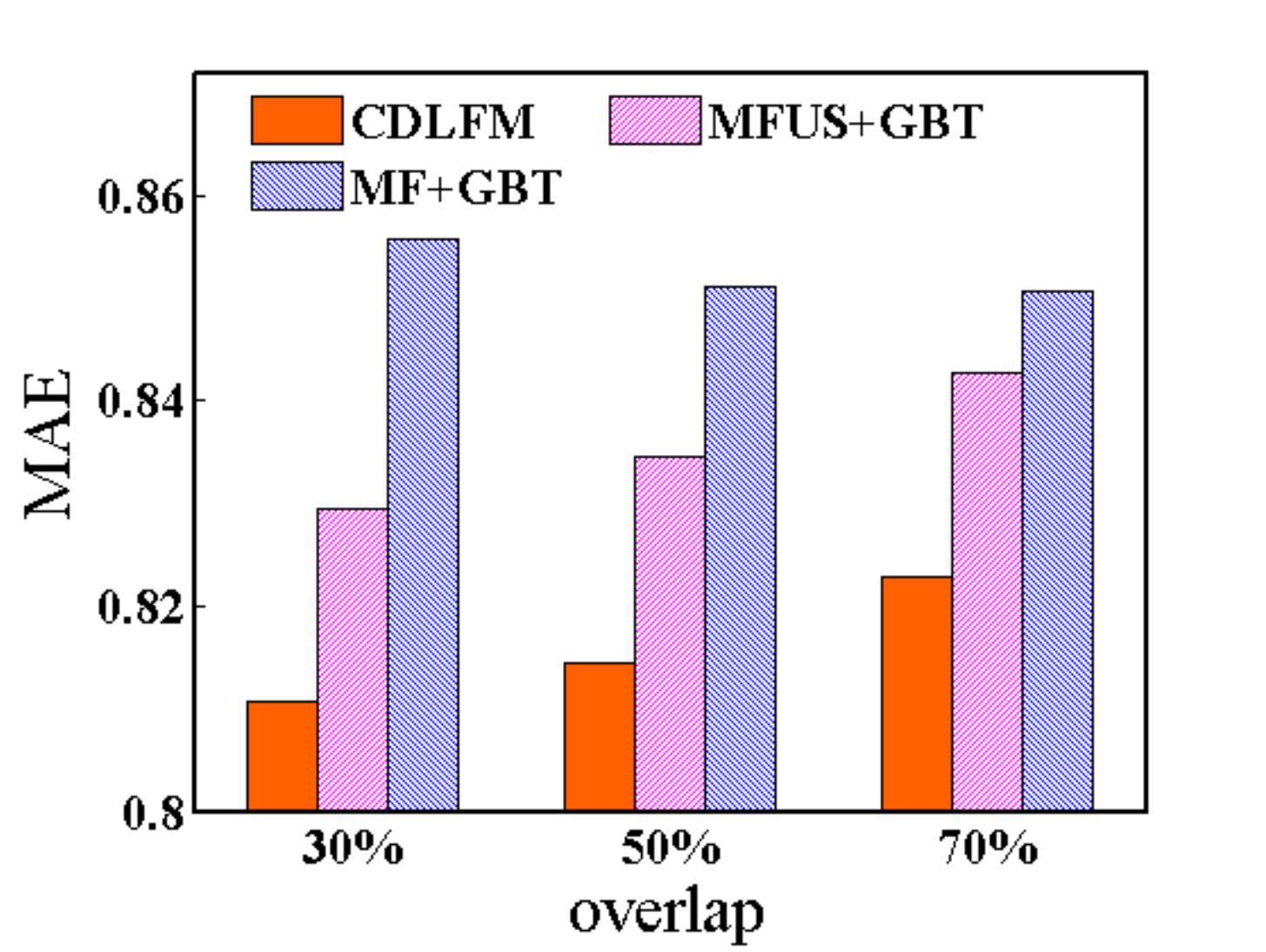}}
\subfigure[MB]{\includegraphics[height=2.99cm, width=2.99cm, angle=0]{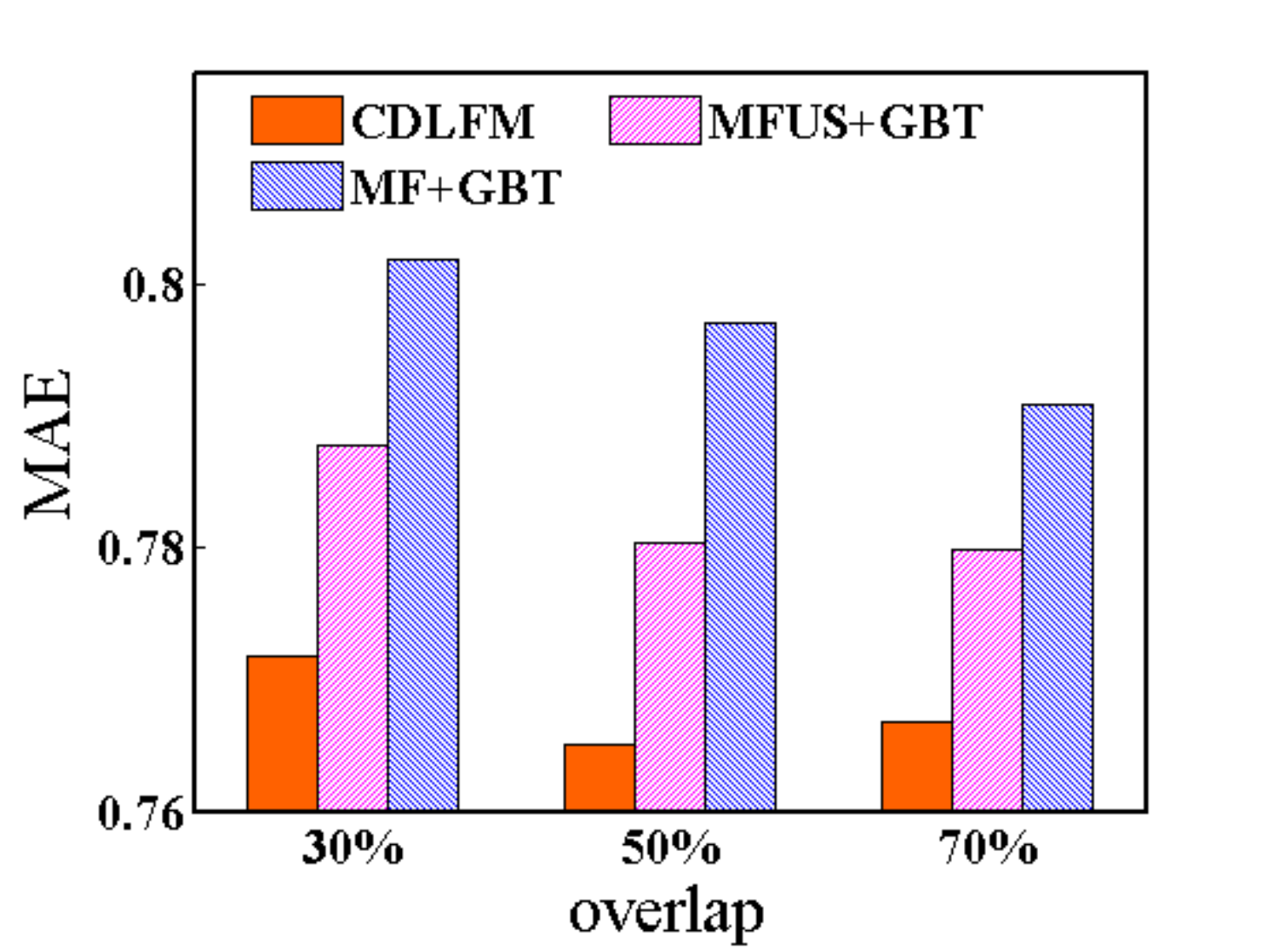}}
\caption{MAE of GBT based methods on Dataset 1}
\label{f7}
\end{figure}

In order to better understand our model, we also implement the methods MF+GBT and MFUS+GBT, which both use the latent feature pairs of all the linked users to learn the feature mapping function. The experimental results of MAE on Dataset 1 are shown in Fig. \ref{f7}. As we can see, MFUS+GBT is better than MF+GBT while CDLFM achieves the best performance, which means the latent features learned from MFUS can improve the performance of GBT and the mapping function learned from similar users are more reasonable. Compared with TMatrix, one can see that GBT is more effective for high-order cross-domain latent features mapping.
\subsection{Parameter Sensitivity Analysis}
Firstly, we study the performance of MFUS on single-domain rating prediction with the movie ratings in Dataset 1. We randomly select 80$\%$ rating data as the training set and the remaining 20$\%$ are used as the test set. We use Backtracking Line Search to accelerate the gradient descent and set $\gamma_{1}=1/4$, $\gamma_{2}=3$, $\gamma_{3}=2$, $\sigma=6$. When $\rho_{1}=0.4$, $\rho_{2}=0.3$, $\rho_{3}=0.3$, the performance of MFUS with different $K$, $\alpha$ and $\beta$ are reported in Table \ref{t2}. One can see that the performance of MFUS first increases and then decreases with the increase of $\beta$. Note that MFUS degrades to MF if $\beta=0$, thus MFUS is better than MF in single-domain rating prediction. For MF, the performance increases with the increase of $K$. However, when $\beta\neq0$, the performance of MFUS becomes worse when $K$ is too large. The reason may be that by considering users' rating behaviors, MFUS can capture more information and low-dimension latent features can characterise users and items accurately. We also conduct experiments to study the influences of aforementioned three similarity measures on MFUS. Table \ref{t3} reports the results with $K$=20, $\alpha=0.01$, $\beta=0.005$. One can observe that with different similarity weights, the prediction performance varies greatly. When $\rho_{1}=0.6$, $\rho_{2}=0.2$, $\rho_{3}=0.2$, we obtain the optimal performance, which implies that the three similarity measures are all useful for an accurate rating prediction.
\begin{table}[t]
\footnotesize
\centering
\caption{Performance of MFUS with varying $K$, $\alpha$ and $\beta$. Numbers in boldface are the best results.}
\begin{tabular}{|p{1.2cm}|p{1.65cm}|p{1.2cm}|p{1.2cm}|p{1.2cm}|p{1.2cm}|p{1.2cm}|p{1.2cm}|}
\hline
\multicolumn{2}{|c|}{\multirow{2}{*}{}} & \multicolumn{3}{c|}{RMSE} & \multicolumn{3}{c|}{MAE} \\
\cline{3-8}
\multicolumn{2}{|c|}{} & $K=15$ & $K=20$ & $K=25$ & $K=15$ & $K=20$ & $K=25$\\ 
\hline
\multirow{5}{*}{$\alpha=0.01$} & $\,\beta=0$ & 1.3301 & 1.308 & 1.2879 & 0.9862 & 0.973 & 0.9611 \\
\cline{2-8}
& $\,\beta=0.001$ & 1.0759 & 1.0709 & 1.0703 & 0.7856 & 0.7837 & 0.7853\\
\cline{2-8}
& $\,\beta=0.002$ & $\textbf{0.9778}$ & $\textbf{0.9796}$ & 1.0364 & $\textbf{0.7309}$ & $\textbf{0.7298}$ & 0.7638\\
\cline{2-8}
& $\,\beta=0.005$ & 0.9791 & 0.9802 & $\textbf{0.9808}$ & 0.7366 & 0.737 & $\textbf{0.737}$\\
\cline{2-8}
& $\,\beta=0.01$ & 0.9913 & 0.9932 & 0.9945 & 0.7499 & 0.7515 & 0.7525\\
\hline
\multirow{5}{*}{{$\alpha=0.1$}} & $\beta=0$ & 1.3064 & 1.2847 & 1.2801 & 0.9692 & 0.955 & 0.9551\\
\cline{2-8}
& $\beta=0.01$ & $\textbf{1.0044}$ & $\textbf{1.0231}$ & $\textbf{1.0519}$ & $\textbf{0.7769}$ & $\textbf{0.763}$ & $\textbf{0.7712}$\\
\cline{2-8}
& $\beta=0.02$ & 1.0194 & 1.0257 & 1.0575 & 0.7948 & 0.7675 & 0.7747\\
\cline{2-8}
& $\beta=0.05$ & 1.0499 & 1.0536 & 1.0646 & 0.8281 & 0.7891 & 0.7857\\
\cline{2-8}
& $\beta=0.1$ & 1.0468 & 1.0673 & 1.0963 & 0.8067 & 0.8024 & 0.8075\\
\hline
\end{tabular}
\label{t2}
\end{table}
\begin{table}[t]
\footnotesize
\caption{Performance of MFUS with varying similarity weights}
\begin{tabular}{|p{0.78cm}|p{0.82cm}|p{0.82cm}|p{0.82cm}|p{0.82cm}|p{0.82cm}|p{0.82cm}"p{0.82cm}|p{0.82cm}|p{0.82cm}|p{0.82cm}|p{0.82cm}|p{0.82cm}|}
\hline
  & \multicolumn{6}{c"}{RMSE} &  \multicolumn{6}{c|}{MAE}\\
\hline
\diagbox[width=0.88cm, height=0.5cm]{$\rho_{1}$}{$\rho_{2}$} & 0 & 0.2 & 0.4 & 0.6 & 0.8 & 1 & 0 & 0.2 & 0.4 & 0.6 & 0.8 & 1\\
\hline
0 & 1.036 & 1.032 & 1.029 & 1.027 & 1.027 & 1.015 & 0.773 & 0.769 & 0.767 & 0.766 & 0.765 & 0.759 \\
\hline
0.2 & 1.043 & 0.978 & 0.984 & 0.99 & 1.029 & - & 0.78 & 0.734 & 0.742 & 0.748 & 0.767  & - \\
\hline
0.4 & 1.05 & 0.977 & 0.983 & 1.033 & - & - & 0.788 & 0.733 & 0.741 & 0.772 & - & - \\
\hline
0.6 & 1.06 & $\textbf{0.976}$ & 1.042 & - & - & - & 0.798 & $\textbf{0.731}$ & 0.781 & - & - & - \\  
\hline
0.8 & 1.073 & 1.059 & - & - & - & - & 0.811 & 0.798 & - & - & - & -\\
\hline
1 & 1.095 & - & - & - & - & - & 0.831 & - & - & - & - & - \\
\hline
\end{tabular}
\label{t3}
\end{table}

\begin{figure}[t]
\centering
\includegraphics[height=3.2cm, width=5.5cm, angle=0]{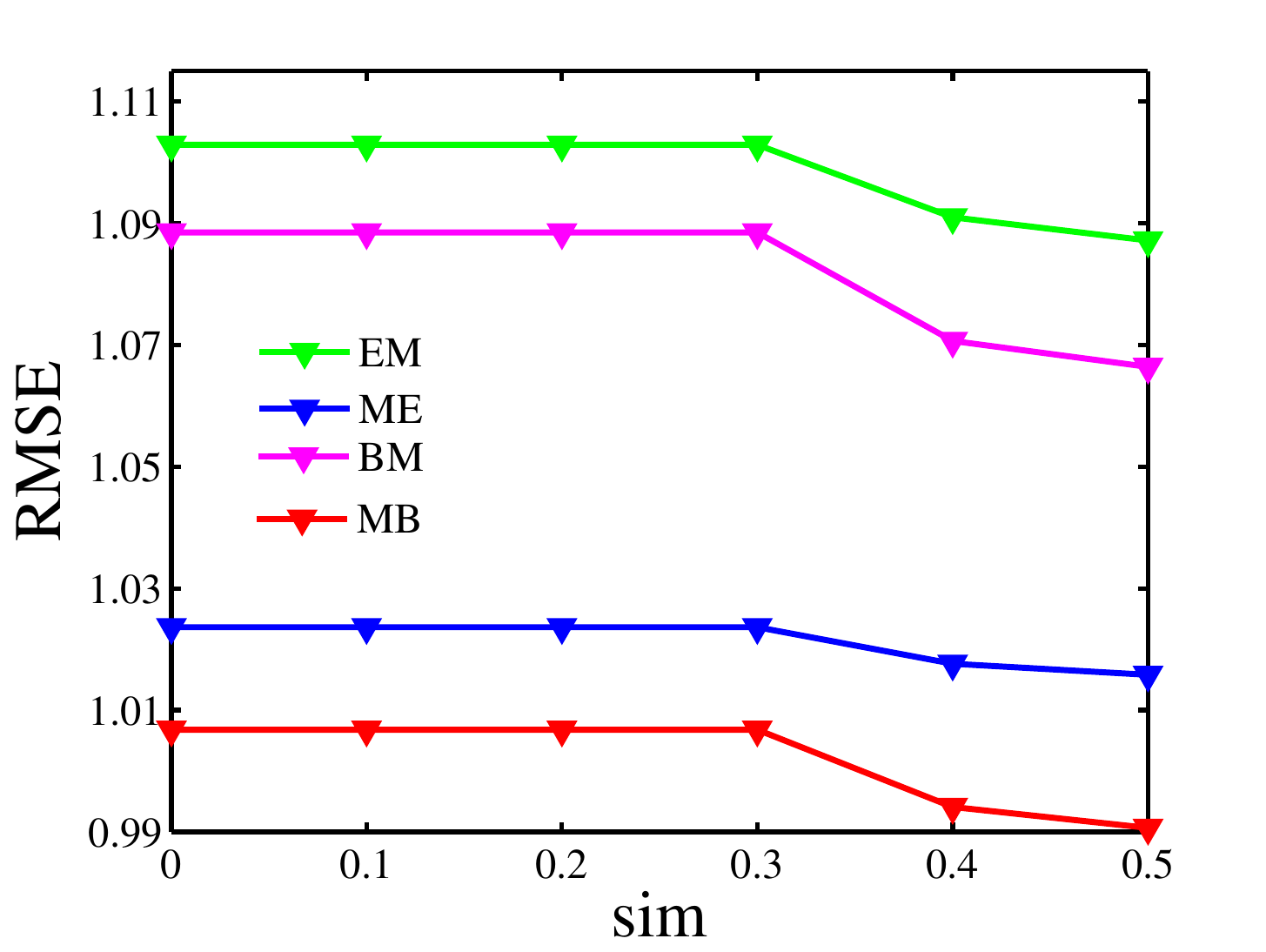}
\includegraphics[height=3.2cm, width=5.5cm, angle=0]{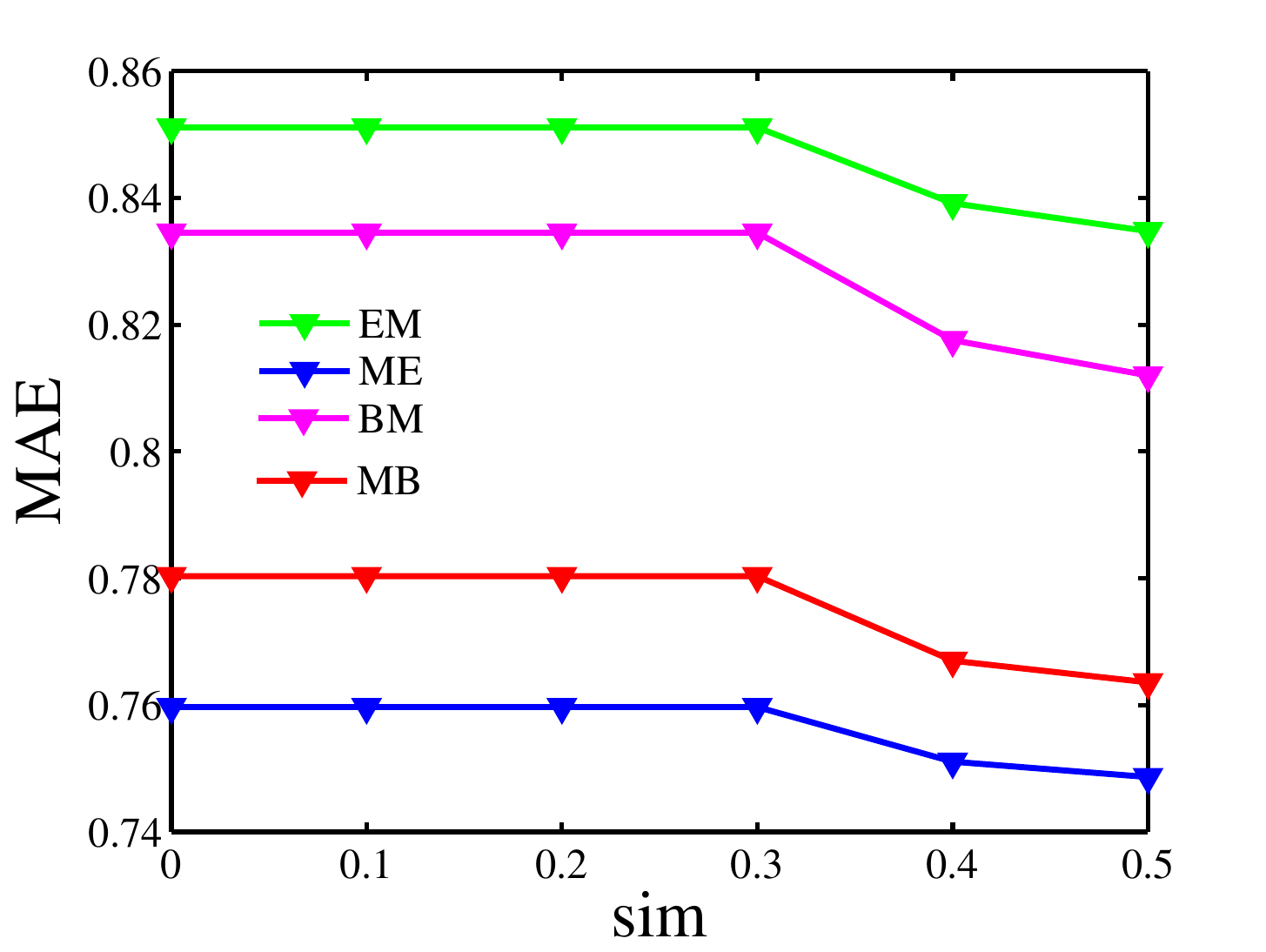}
\caption{Performance of CDLFM on different datasets with different $sim$ values}
\label{f8}
\end{figure}

Next, we study the effect of the parameter $sim$ to our CDLFM model. We conduct experiments with user overlap level 50$\%$ and the results are reported in Fig. \ref{f8}. Because the computed similarities are all larger than 0.2 and some users have no neighbor linked users when $sim=0.5$, the lines are flat at first and the largest studied value of $sim$ is 0.5. One can see, with a larger $sim$, we can learn more accurate feature mapping functions. The results demonstrate that for a group of users who have similar rating behaviors in an item domain, they tend to be consistent in certain aspects in another item domain.
\section{Conclusions}
In this paper, we present a novel model CDLFM for more effective cross-domain recommendation for cold-start users. Firstly, we propose a new rating matrix factorization model by incorporating user similarities, by which we can take users' rating behaviors into consideration and learn more accurate latent features of users in sparse domains. Then, we propose a neighborhood based GBT method to learn the high-order latent feature mapping function across domains. The experimental results show that our model outperforms other state-of-the-art methods on the problem of cross-domain recommendation for cold-start users.
\subsubsection{Acknowledgements.}
This work is supported by NSF of China (No. 61602237, No. 61672313), 973 Program (No. 2015CB352501), NSF of Shandong, China (No. ZR2017MF065), NSF of Jiangsu, China (No. BK20171420). This work is also supported by US NSF through grants IIS-1526499, and CNS-1626432.

\end{document}